\documentclass[review,onefignum,onetabnum]{siamart190516}
\usepackage{amsfonts,amssymb}
\usepackage{amsmath}
\usepackage{graphicx,color,adjustbox}
\usepackage{subfigure}
\usepackage{graphicx}
\usepackage{dcolumn}
\usepackage{bm}
\newtheorem{thm}{Theorem}
\newtheorem{lem}[thm]{Lemma}
\newtheorem{prop}[thm]{Proposition}
\newtheorem{examp}{Example}

\definecolor{bgreen}{rgb}{0.0, 0.7, 0.0}

\definecolor{cmarkgreen}{rgb}{0.0, 0.5, 0.0}
\definecolor{xmarkred}{rgb}{0.8, 0.0, 0.0}

\definecolor{ballblue}{rgb}{0.63, 0.79, 0.95}
\definecolor{bgreen}{rgb}{0.0, 0.7, 0.0}
\nolinenumbers
\title{Reconstruction of network dynamics \\ from partial observations}

\author{Tyrus Berry and Timothy Sauer\thanks{Department of Mathematical Sciences, George Mason University, Fairfax, VA (\email{tsauer@gmu.edu})}}

\begin{document}

\maketitle
\begin{abstract}
We investigate the reconstruction of time series from dynamical networks that are partially observed. In particular, we address the extent to which the time series at a node of the network can be successfully reconstructed when measuring from another node, or subset of nodes, corrupted by observational noise. We will assume the dynamical equations of the network are known, and that the dynamics are not necessarily low-dimensional. The case of linear dynamics is treated first, and leads to a definition of observation error magnification factor (OEMF) that measures the magnification of noise in the reconstruction process.  Subsequently, the definition is applied to nonlinear and chaotic dynamics. Comparison of OEMF for different target/observer combinations can lead to better understanding of how to optimally observe a network. As part of the study, a computational method for reconstructing time series from partial observations is presented and analyzed.
\end{abstract}


\section{Introduction}

The subject of network dynamics is increasingly relevant to physical process modeling. Networks present a fascinating departure from generic dynamical systems due to the constraints imposed on direct communication between nodes, resulting in complicated dynamics and nontrivial bifurcation structures \cite{golubitsky2006nonlinear,sauer2004reconstruction,boccaletti2006complex,newman2011structure}.  Modeling by networks has become an important topic in almost every area of physical and biological science, including distributed mechanical processes, weather and climate, and metabolic, genomic and neural networks.

An important aspect of understanding distributed systems is the choice of observables that facilitate reconstruction of the collective dynamics of the network. The theory of observability was pioneered for linear dynamics by Kalman \cite{kalman1959general}. For nonlinear dynamics, the theory of attractor reconstruction \cite{Takens,SYC} provides hope that for generic observables of sufficiently high dimension, the dynamics can be reconstructed. Although observations at single or even multiple nodes of a network may not be provably generic, 
some aspects of reconstructibility may  be present by observing even a single node in a strongly connected network, i.e. a network for which every node is downstream from every other node. 
The ability to reconstruct dynamics from partial observations in both linear and nonlinear networks is a topic of intense recent interest, with close connections to questions of observability \cite{letellier2005relation,letellier2005graphical,letellier2009symbolic,liu2013observability,sendina2016observability,haber2017state,sendina2022observability,whalen2015observability,montanari2022functional} and controllability  \cite{lin1974structural,liu2011controllability,wang2014controllability}.  

However, even in the linear case, observability in theory does not guarantee a satisfactory reconstruction in practice, in particular from data collected from a sparsely--connected network, or far from target nodes, even in the  case where the equations of motion are known. To date, even in this more tractable scenario, surprisingly little in the way of general practical requirements  have been developed for inferring information from measurements. A critical obstruction is the presence of noise in the observations, and the tendency of noise to be magnified in efforts to reconstruct the dynamics. In this article, we analyze a definition of error magnification in reconstruction of network trajectories,  first introduced in \cite{guan2018limits}, and exhibit its behavior for some relevant examples. The main conclusion is that for practical use of network trajectory reconstruction techniques, theoretical observability may be only a first step, and that a multiplier  that measures error magnification, akin to condition number in matrix calculations, may fundamentally govern the limits of reconstructibility. In short, if the noise level at the observer is $\sigma$ times the macroscopic variability of the dynamics, then the error magnification must be on the order of $1/\sigma$ or lower to allow accuracy in the reconstructed dynamics. Our first aim is to quantify this magnification for each specific observing subset and target node of the network.

A second objective of this article is to present a general method to infer the time series of  the entire network using only observations from a fixed subset of the nodes. This objective, the reconstruction of time series from partial observations, is often called "spin-up" in data assimilation problems.  As used in modern applications such as weather forecasting, the spin-up phase is equivalent to developing a consistent set of initial conditions that the atmosphere, for example, obeys. This initial condition, along with the known equations, allows the ``digital twin'' formed by data assimilation techniques to mirror the real system. 

Our focus on these two problems is due to the obvious intrinsic utility of not only being able to reconstruct unmeasured dynamics at nodes by measuring other nodes, but also to understand the relative difficulty levels of different potential reconstructions. Such considerations allow the user to decide where to efficiently observe the network if choices are available. 

Because of the extensive published work on related problems, it is important to clarify our goals in this article. The term ``network reconstruction'' often refers to determination of the network structure: Inferring unknown structural elements of a network from observations, either ab initio or under certain known constraints. In this article, we are attacking a different problem, which we will call ``network trajectory reconstruction''. In this scenario, the network, including all network connections and relations between nodes is known, and we are seeking to reconstruct the trajectory. As mentioned above, in cases of partial observation, network reconstruction alone does not complete the analysis of a system, because the ability to reconstruct network trajectories is highly dependent on where the observations are made.  Even in a completely known, strongly connected network, we will find that the reconstruction error depends in a complicated way on the location of the observations. Therefore, awareness of the error magnification that occurs between observation and reconstruction is crucial for the modeler. There is a close analogy to the question in numerical linear algebra, even when a matrix is completely known, about how errors are magnified from the input data to the output solution. The influential work of Wilkinson \cite{wilkinson1963rounding}, based on a suggestion of Turing \cite{10.1093/qjmam/1.1.287}, popularized the key concept of condition number to shed light on this issue. The error magnification factor developed below is a network reconstruction analogue to this valuable concept.

Thus we are not directly addressing many other pertinent questions about dynamical networks, for example: (1) reconstructing dynamics at other network nodes when observations of the full network have been previously made, or (2) reconstructing the network  or network equations themselves from full or partial observations. Problem (1) is connected with Takens' theorem and its analogues, and is unlikely to be effective outside the domain of low-dimensional dynamics. In contrast, no dimensional restrictions are necessary in the current work.  Problem (2) is also under intense development  but is not the subject of this paper. 

We begin in Section 2 with a review of observability in the discrete linear case, presented in a way to simplify our later discussion of error magnification. In Section 3 we define the Observation Error Magnification Factor, and in Section 4 we present a numerical method for reconstruction for nonlinear time series. Section 5 contains results of applying the methods to nonlinear networks.

\section{Discrete linear networks}

In this section, we collect some elementary principles of observing networks in the simplest case of linear dynamics. Most of these ideas date back at least to Kalman \cite{kalman1959general}.  Our description is designed to lead to a practical notion of observational error magnification that we can later extend to the nonlinear case.

Consider the discrete  linear dynamical system $x^{k+1} = Ax^k$ on $R^n$ where $A$ is an $n\times n$ matrix. We can write down the relation between the initial state $x^0 = [x^0_1, x^0_2, x^0_3 ,\ldots, x^0_n]$ and the time series observed at an arbitrary node $j$, which is denoted by $s_k= (A^kx^0)_j$ for $k = 0. 1. 2. \ldots$.
For $t>0$, consider the matrix 
\begin{equation} \label{e5}
M_{t,j} = 
\left[
\begin{array}{cccccc}
0& \cdots& 0& 1& \cdots&0\\ 
 (A)_{j1}&\cdots&(A)_{j,j-1}&(A)_{jj}&\cdots&(A)_{jn}\\
(A^2)_{j1}&\cdots&(A^2)_{j,j-1}&(A^2)_{jj}&\cdots&(A^2)_{jn}\\
\vdots&&\vdots&\vdots&&\vdots\\
(A^{t-1})_{j1}&\cdots&(A^{t-1})_{j,j-1}&(A^{t-1})_{jj}&\cdots&(A^{t-1})_{jn}
\end{array}
\right] = 
\left[
\begin{array}{c}
(A^0)_j\\
(A^1)_j\\
\vdots\\
(A^{t-1})_j
\end{array}
\right] 
\end{equation}
That is, the rows of $M_{t,j}$ consist of the $j$th rows of $A^k$ for $0\le k <t$. Note that
\begin{equation}M_{t,j} 
\left[
\begin{array}{c}
x^0_1\\ x^0_2\\ x^0_3\\ \vdots\\ x^0_n
\end{array}\right]
 = \left[
\begin{array}{c}
s_0\\ s_1\\ s_2\\ \vdots \\s_{t-1}
\end{array}\right],  \label{e1}
\end{equation}
which is the connection between initial conditions and observations that we will be able to exploit. First, we define the concept of a kernel node.

\bigskip

\noindent {\bf Defn.} Let $M$ be an $m\times n$ matrix. Call node $i$ (or variable $i$) a {\bf kernel node} (or variable) for  $M$ if $v_i\neq 0$ for some vector $v = [v_1,\ldots, v_n]$ in the nullspace of $M$, and a {\bf regular node} otherwise. More generally, let $M$ be a structured matrix, i.e. one with only certain specified nonzero entries. We call node $i$ a {\bf generic kernel node} for $M$ if for Lebesgue-almost every choice of specified nonzero entries in $M$, respecting the structure of $M$, there is $v\in\ $null$(M)$ such that $v_i\neq 0$; we call $i$ a {\bf generic regular node} if $v_i=0$ for every vector $v\in\ $null$(M)$ for almost every choice of entries in $M$. 

By definition, for a specific matrix $M$, every node is either a kernel node or a regular node. Interestingly, the same is true for nodes of a generic linear network.

\bigskip

\noindent{\bf Fact 1.} \cite{jahedi2023structured} For matrices $M$ with a given structure, every node is either a generic kernel node or a generic regular node. 

\bigskip

 If $M_{t,j}$ cannot be made to be full rank by taking $t$ sufficiently large, there are distinct initial states which result in the same observations at the j-th node and thus produce  indistinguishable time series.  The vectors in the kernel of $M_{t,j}$ will have non-zero entries for nodes that are related to the indistinguishable states.  Conversely, if every vector in the kernel of $M_{t,j}$ has a zero entry for the $i$th node, this indicates that the $i$th node does not participate in any of the indistinguishable states, and so the state of the $i$th node will be identifiable from observations at the $j$th node. These properties are summarized in the following proposition.

\begin{prop} \label{p1}
Consider the observations $s_k = (A^kx^0)_j$ at node $j$. 
The following are equivalent:\\ (1) The initial state of node $i$ is uniquely determined by infinite sequence $\{s_0,s_1,\ldots\}$.\\ (2) The initial state of node $i$ is uniquely determined by finite sequence $\{s_0,\ldots,s_{n-1}\}$. (3) Node $i$ is a regular node of $M_{n,j}$. \end{prop} 

\noindent{\bf Proof.} By the Cayley-Hamilton Theorem, the matrix A satisfies its own characteristic equation, so the row vectors $e_j^TA^0,...,e_j^TA^n$ are linearly dependent, where $e_j$ denotes the $j$th column of the identity matrix. This implies that
rows $n+1, n+2, \ldots, t$ of $M_{t,j}$ are linear combinations of the rows of $M_{n,j}$. Therefore for $t \ge n$, $v\in$ ker $M_{t,j}$ if and only if $v\in$ ker $M_{n,j}$, and it reduces to proving (2) and (3) are equivalent.

If $v\in$ ker $M_{n,j}$ and $v_i \neq 0$, then for any solution $x=[x_1,\ldots, x_n]$ of (\ref{e1}), $x+\alpha v$ are also solutions and the $i$th coordinates have many different values, contradicting uniqueness. Conversely, if there are solutions with two different values at the $i$th coordinate, then there is a kernel node with nonzero $i$th coordinate. \qed

\begin{figure}
 \begin{center}
\subfigure[] {\includegraphics[width=.25\textwidth]{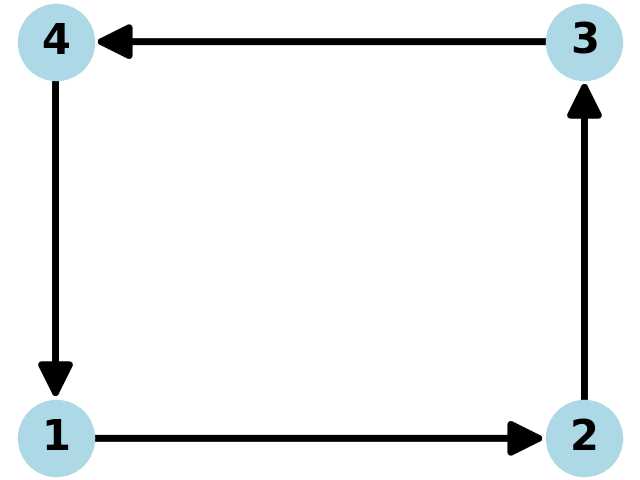}}\hspace{.1\textwidth}
\subfigure[]{\includegraphics[width=.25\textwidth]{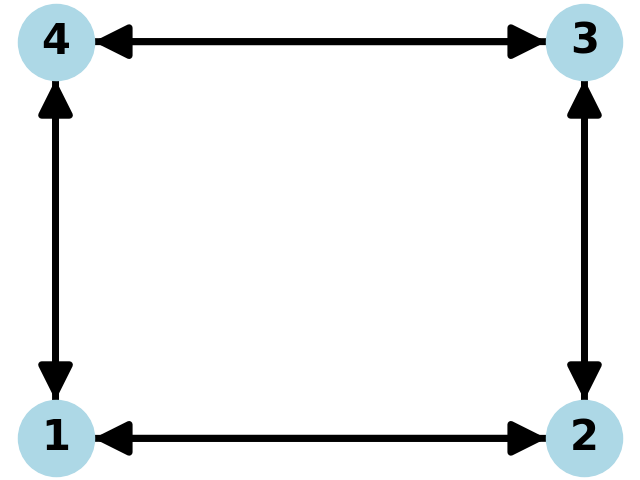}}\hspace{.1\textwidth}
\subfigure[]{\includegraphics[width=.25\textwidth]{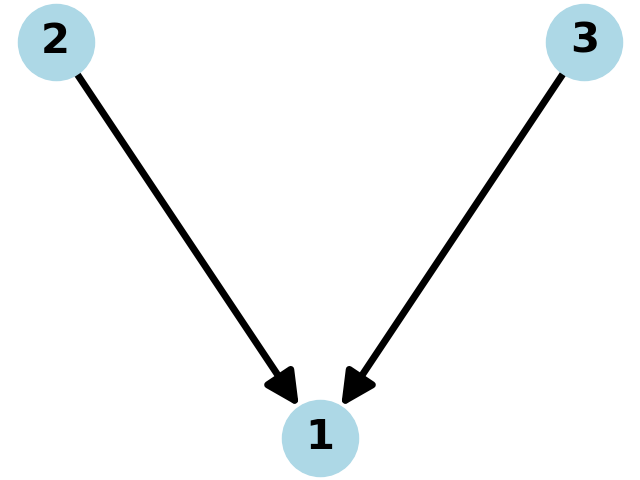}}
\end{center}
\caption{\rm Three example networks. (a) A 4-node network that is reconstructible from any node for generic weights. (b) An undirected network, also reconstructible from any node.  (c) Three-node network that exhibits a bottleneck for generic weights, which obstructs the reconstruction of nodes 2 or 3 from observations at node 1. } \label{f1}
\end{figure}

\begin{examp} \label{ex1} \rm Fig.~\ref{f1}(a) shows a 4-node network that is fully reconstructible, in the sense that generically, measurements at any node can be reconstructed from observations at any other node. The edge weight matrix $A$ is
$$A = \left[
\begin{array}{cccc}
0 & 0 & 0&a_{14}\\
a_{21}&0&0&0\\
0&a_{32} &0&0\\ 
0&0&a_{43}&0
\end{array}\right]$$
where the $a_{ij}$ are arbitrary entries into this structured matrix, and the rest of the entries are fixed at zero. Without loss of generality, consider observing at node $j=1$, for which the observation matrix  is
$$M_{4,1} =
\left[
\begin{array}{cccc}
1&0&0&0\\
0&0&0&a_{14}\\
0&0&a_{14}a_{43}&0\\
0&a_{14}a_{43}a_{32}&0&0
\end{array}\right].
$$
The determinant of the matrix is $a_{14}^3a_{43}^2a_{32}$. It follows that for generic entries (avoiding the lower-dimensional hypersurface defined by the determinant) there are no kernel nodes, so according to Proposition \ref{p1}, we will be able to reconstruct each node from the time series at node 1.
\end{examp}

\begin{examp} \rm Fig.~\ref{f1}(b) shows a 4-node network that is also fully reconstructible, in the same way as Example \ref{ex1}. The edge weight matrix $A$ is
$$A = \left[
\begin{array}{cccc}
0 & a_{12} &0& a_{14}\\
a_{21}&0&a_{23}&0\\
0&a_{32} &0&a_{34}\\ 
a_{41}&0&a_{43}&0
\end{array}\right]$$
where the $a_{ij}$ are arbitrary entries into this structured matrix, and the rest of the entries are fixed at zero. Without loss of generality, consider observing at node $j=1$, for which the observation matrix  is
$$ M_{4,1}
 = \left[
\begin{array}{cccc}
1&0&0&0\\
0&a_{12}&0&a_{14}\\
a_{12}a_{21}+ a_{14}a_{41}&0& a_{12}a_{23} + a_{14}a_{43}&0\\
0&m_{42}&0&m_{44}
\end{array}\right].
$$
where $m_{42} = a_{12}(a_{12}a_{21} + a_{14}a_{41}) + a_{32}(a_{12}a_{23} + a_{14}a_{43}) $ and
$m_{44} = a_{14}(a_{12}a_{21} + a_{14}a_{41}) + a_{34}(a_{12}a_{23} + a_{14}a_{43}) $.
The determinant of the matrix is a (nonzero) degree 6 polynomial in the $a_{ij}$. For generic entries (avoiding the lower-dimensional hypersurface defined by the determinant) there are no kernel nodes, so according to Proposition \ref{p1}, the time series at each node is reconstructible from the time series at node 1.
\end{examp}

Next we consider how the topology of the network affects reconstructibility from a node. One might expect, for example, that trajectories of all upstream nodes can be successfully reconstructed from a downstream node. We say that a node $i$ is upstream from node $j$ if  there is a path through the directed network from a node $i$ to the node $j$. Can the trajectories at node $i$ be reconstructed from the time series observed at node $j$? The answer is no, even if we assume the network weights are generic.

\begin{examp} \rm
 A simple instance is shown in Fig.~\ref{f1}(c). If we assume the weights of the graph are $a_{12}$ and $a_{13}$ on the left and right arrows, respectively, then the identically zero time series observed at node 1 can be explained by zeros on nodes 2 and 3, or alternatively by any multiple of the constant time series $(-a_{13}, a_{12})$. Therefore the time series at nodes 2 and 3 are not uniquely determined by measurements at node 1.

This is an illustration of Proposition \ref{p1}, since
$$A = \left[
\begin{array}{ccc}
0 & a_{12} & a_{13}\\
0&0&0\\0&0&0
\end{array}\right],\ \ \ \ \ 
M_{3,1} = \left[
\begin{array}{ccc}
1&0&0\\
0 & a_{12} & a_{13}\\
0&0&0
\end{array}\right],
$$
so nodes 2 and 3 are identified as kernel nodes for $M_{3,1}$, no matter what the values $a_{12}$ and $a_{13}$ are. 
\end{examp}

\begin{figure}
 \begin{center}
 \subfigure[]{ \includegraphics[width=.4\textwidth]{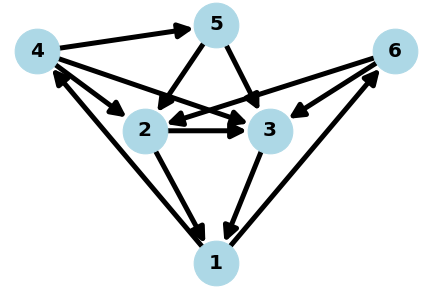}}\hspace{.1\textwidth}
 \subfigure[]{ \includegraphics[width=.4\textwidth]{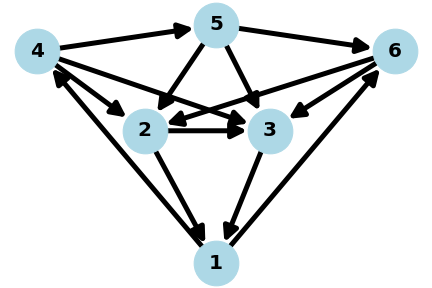}}
\end{center}
\caption{\rm (a) Example of 6-node network with a $1$-bottleneck consisting of the subset $B=\{2, 3, 5, 6\}$. (b) Adding one edge from node 5 to node 6 eliminates the bottleneck } \label{f2}
\end{figure}

\begin{examp} \rm
Consider the graph in Figure \ref{f2}(a). Its edge weight matrix is 
$$A = \left[
\begin{array}{cccccc}
0& a_{12}& a_{13}&   0&   0&   0\\
  0&   0&   0& a_{24}& a_{25}& a_{26}\\
0& a_{32}&   0& a_{34}& a_{35}& a_{36}\\
a_{41}&   0&   0&   0&   0&   0\\ 
  0&   0&   0& a_{54}&   0&   0\\
a_{61}&   0&   0&   0&   0&   0
\end{array}\right].$$
We can see that all nodes are upstream from node 1. However, it turns out that while the time series at node 4 can be successfully reconstructed from the time series at node 1, it can verified (somewhat laboriously with symbolic algebra) that nodes 2, 3, 5, and 6 are kernel nodes, and  by Proposition \ref{p1}, cannot be uniquely reconstructed from the time series at node $1$. In fact, the concept of bottlenecks, introduced by Lin in \cite{lin1974structural} and further developed in \cite{jahedi2023structured}, can make it much easier to diagnose obstructions like this one directly from the topology of the graph.  
\end{examp}

Let $S$ be a subset of nodes of a directed graph with $n$ nodes. We will denote by $S^\to$ the {\bf forward set} of $S$, the set of all nodes such that there is an arrow to the node from a node in $S$. We call the sets $\{S,S^\to\}$ a {\bf k-bottleneck} if $|S| = |S^\to| + k$ for some $k>0$. A {\bf minimax} $k$-bottleneck is a $k$-bottleneck of maximal $k$ that is minimal with the property of being a $k$-bottleneck, more precisely, such that no subset is a $k$-bottleneck. 

It turns out that there is a
connection between minimax bottlenecks and the nullspace of $A$. The following fact is proved in  \cite{jahedi2023structured}:

\bigskip

\noindent{\bf Fact 2.} For generic entries respecting the structure of $A$, rank$(A)< n$ if and only if there is a $k$-bottleneck for some $k>0$. Moreover, the minimax bottleneck consists of the generic kernel nodes of $A$. 

\bigskip

For example, consider the set $S=\{2,3,5,6\}$ in Fig. \ref{f2}(a). Note that $S^\to = \{1, 2, 3\}$. This is a minimax $1$-bottleneck, so the nodes 2, 3, 5, and 6 are generic kernel nodes of the structured matrix $A$.

The following key observation shows that kernel nodes for the weight matrix $A$ are also kernel nodes for the observation matrix $M_{n,j}$.
\begin{proposition} \label{p2}
Let $A$ be a structured matrix and let $S$ = set of generic kernel nodes of $A$. If $j\notin S$, the set $S$ consists of generic kernel nodes for $M_{n,j}$.
\end{proposition}

\noindent {\bf Proof.} Let $v \in\ $ null$(A)$. Since $j\notin S$, the first row of $M_{n,j}$ is orthogonal to $v$. Since $Av = 0, A^2v = 0, \ldots$, each row of $M_{n,j}$ after the first is also orthogonal to $v$. Thus $v\in\ $null$(M_{n,j})$.

\bigskip

By Proposition \ref{p2}, the time series at 2, 3, 5 and 6 cannot be reconstructed from node 1 or from node 4. This is not surprising for node $j=4$, since not all nodes are upstream from 4. However, it is interesting for $j=1$, because all nodes are upstream from node 1.
One can further check using Proposition 1 that the time series at node 4 can be uniquely reconstructed from observing at node 1, even though the rest of the nodes cannot.

Knowledge of the bottleneck, $S=\{2,3,5,6\}\rightarrow S^\to =\{1, 2, 3\}$  in this case, shows what is necessary to remove it. By connecting nodes 5 and 6 as in Fig.~\ref{f2}(b), $S^\to$ becomes $\{1, 2, 3, 6\}$, destroying the bottleneck. In the revised network, all nodes are reconstructible from node 1.

\bigskip

\begin{figure}
 \begin{center}
 \subfigure[]{ \includegraphics[width=.4\textwidth]{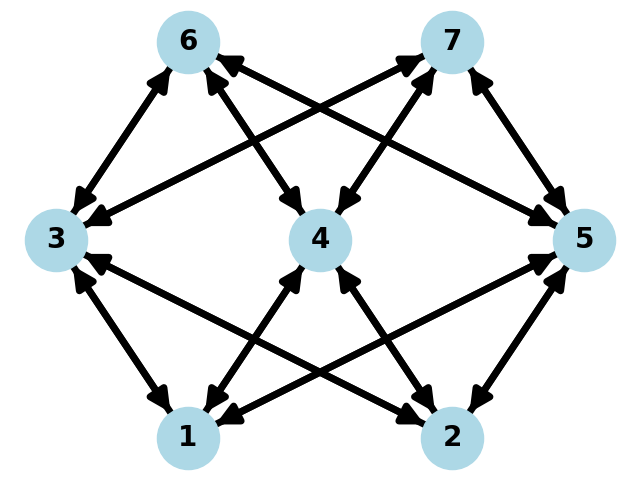}}\hspace{.1\textwidth}
 \subfigure[]{ \includegraphics[width=.4\textwidth]{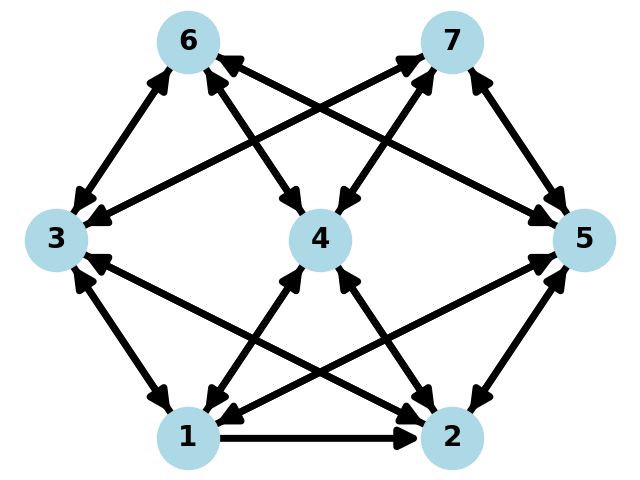}}
\end{center}
\caption{\rm Example of 7-node network. (a) The graph has a $1$-bottleneck with $S =  \{1, 2, 6, 7\}$, which obstructs reconstructibility. (b) Adding one extra directed edge to the graph allows the time series at every node to be reconstructed from every other node.} \label{f3}
\end{figure}

\begin{examp}\rm
The graph in Fig.~\ref{f3}(a) is a further illustration of a bottleneck. Note that $S=\{ 1, 2, 6, 7\} \to S^\to = \{3, 4, 5 \}$ is a 1-bottleneck, so by Fact 2, nodes 1, 2, 6 and 7 are generic kernel nodes for $A$, and according to Proposition \ref{p2}, also  for $M_{n,j}$ for $j = 3, 4$ and $5$.  Therefore nodes 3, 4, and 5 cannot be used to reconstruct time series from the other nodes. 

As in the previous example, if we could add another edge, say from node 1 to node 2, the bottleneck disappears. Therefore the network in Fig.~\ref{f3}(b) is reconstructible from observations at any node. However, while this is theoretically true, there may be a price to pay in such a marginal case. We return to this network shortly in Example \ref{ex7} to examine how practical this would be.
\end{examp}

Although we have focused on obstructions to reconstruction up to this point, it is still likely that wide swaths of examples are fully reconstructible, even in the discrete linear case, if problems such as bottlenecks are avoided. In particular, we suggest two sets of hypotheses that preclude bottlenecks and can provide lots of successful examples of reconstructing times series from partial observations. For a fixed directed graph with $n$ nodes, denote by $A$ the structured matrix of edge weights $a_{ij}$. A subset of nodes in a directed graph is {\bf strongly connected} if there is a directed path connected from node $i$ to node $j$ for any pair of nodes $i, j$ in the subset.

{\bf Conjecture 1.}
Assume that all nodes have self-connections, i.e. assume $a_{ii}\neq 0$ for $1\leq i \leq n$. Then for almost every choice of entries of the edge weight $A$, each node upstream of node $j$ can be observed from node $j$. More precisely, if there is a directed path from node $i$ to node $j$,  then the time series of node $i$ can be reconstructed from the time series observed at $j$.

{\bf Conjecture 2.}  For almost every choice of entries $a_{ij}$ in a structured edge weight matrix $A$, any strongly connected subset of nodes can be observed from any node in the subset.

\section{Observational Error Magnification Factor} \label{sec3}

 In the previous section we discussed necessary conditions that imply that a given trajectory at some network node can be uniquely reconstructed from observations at another node. Given that such a reconstruction exists, we next turn to whether it is feasible in an experimental context to carry out the reconstruction. To this end, we investigate the role of noise in the reconstruction. In particular, we will define the Observational Error Magnification Factor, that quantifies the conditioning of the problem of reconstructing one time series from another. As in the previous section, we begin by looking at the linear case. 
 
In particular, assume we want to reconstruct the time series at node $i$ from the observations at node $j$.  We will achieve this by first using $M_{t,j}$ to estimate the initial state of the network from the observations at node $j$, and then applying $M_{t,i}$ to the estimated initial condition in order to reconstruct the observations at node $i$.  The goal is then to determine how the size of a random perturbation to the observation at node $j$ will effect the error in the resulting estimate of the time series at node $i$.
 
Consider initial condition $x^0=[x_1^0, \ldots, x_n^0]$ of the network and let $s=[s_1, \ldots, s_t]$ denote the time series of length $t$ observed (exactly) at some node from which all nodes can be reconstructed. Let node $j$ be the observing node and let $M_{t,j}$ be the matrix of (\ref{e5}), which by our assumption and Propostion \ref{p1} has full rank $n$. According to this equation, $M_{t,j}x^0 = s$. Now assume we observe node $j$ with noise level $\sigma$ and attempt to reconstruct the initial condition $x^0$. Thus we observe $s+e$ for some observational error $e$. The deviation of the initial condition of the true trajectory due to this observational error can be denoted $h^0$, defined by $M_{t,j}(x^0+h^0) = s+e$, where the error $e=[e_0,\ldots,e_t]$ satisfies $\mathbb{E}(e_k) = 0$ and $\mathbb {E}(e_k^2) = \sigma^2$. Since in general $t>n$, it is unlikely that this problem has an exact solution, so we must consider the corresponding least squares problem:
\begin{equation} \label{e6}
\min_{h^0} ||M_{t,j}(x^0+h^0)-(s+e)||_2 = \min_{h^0} ||M_{t,j}h^0-e||_2.
\end{equation}
The minimum-norm least squares solution of (\ref{e6}) is $$h^0 = M_{t,j}^\dagger e,$$
where $M_{t,j}^\dagger$ denotes the pseudoinverse of $M_{t,j}$.  This expression makes sense even if $M_{t,j}$ is not full rank. To compute $M^\dagger$, let $M_{t,j}=USV^T$ be the singular value decomposition of $M_{t,j}$, and set $M_{t,j}^\dagger = VS_{\rm inv} U^T$, where $S_{\rm inv}$ is the diagonal matrix of the same shape as $M_{t,j}$ for which each diagonal entry is the reciprocal of the corresponding entry of $S$ if it is nonzero, and zero otherwise. 

Now that we have an optimal estimate of the initial state (using observations at node $j$), we are ready to reconstruct the time series at node $i$.  We need to apply $M_{t,i}$ from (\ref{e5}) to the least squares initial condition $x^0+h^0$. Since $M_{t,i}x^0$ gives the true time series at node $i$, the length-$t$ time series of perturbations at node $i$ generated by $h^0$ is  
\begin{equation}
h = M_{t,i}h^0 = M_{t,i}M_{t,j}^\dagger e.  
\end{equation}
The error magnification factor will be defined in terms of root-mean-squared (RMS) error.

\bigskip

{\bf Defn.} For a random vector $v\in \mathbb{R}^n$, define RMS($v$) ${\displaystyle  = \left(\frac{1}{n}\mathbb{E}[\sum_{i=1}^n v_i^2]\right)^{1/2}}$

\begin{lem} \label{lm2}
Let $A$ be an $m\times n$ matrix and let $e=[e_1,\ldots,e_n]$ be a random vector with $\mathbb{E}[e_i] = 0, \mathbb{E} [e_i^2] = \sigma^2$, and $\mathbb{E}[e_ie_j] = 0$ for $i\neq j$. Then  
\begin{equation} {\rm RMS}(Ae) = \frac{1}{\sqrt{m}}||A||_F\ {\rm RMS}(e) =\frac{\sigma}{\sqrt{m}}||A||_F,
\end{equation}
 where $||\ ||_F$ denotes the Frobenius norm.
\end{lem}
{\bf Proof.} Note that for $v\in\mathbb{R}^n$ we have $RMS(v)^2 = \mathbb{E}[v^Tv/n] = \mathbb{E}[\textup{trace}(vv^T)/n]$ so  
\begin{align*} \textup{RMS}(Ae)^2 &= \mathbb{E}[\textup{trace}(Aee^TA^T)/m] =\mathbb{E}[\textup{trace}(A^TAee^T)/m] \\
&= \textup{trace}(\mathbb{E}[A^TAee^T])/m = \textup{trace}(A^TA\mathbb{E}[ee^T])/m \\
&= \textup{trace}(A^T A \sigma^2 I)/m = \frac{\sigma^2}{m}||A||_F^2 = \frac{1}{m}||A||_F^2 \textup{RMS}(e)^2
\end{align*}
where we applied invariance of trace to cyclic permutations, linearity of expectation, and $||A||_F^2 = \textup{trace}(A^T A)$.
\qed \bigskip

\begin{figure}
 \begin{center}
 \subfigure[]{ \includegraphics[width=.47\textwidth]{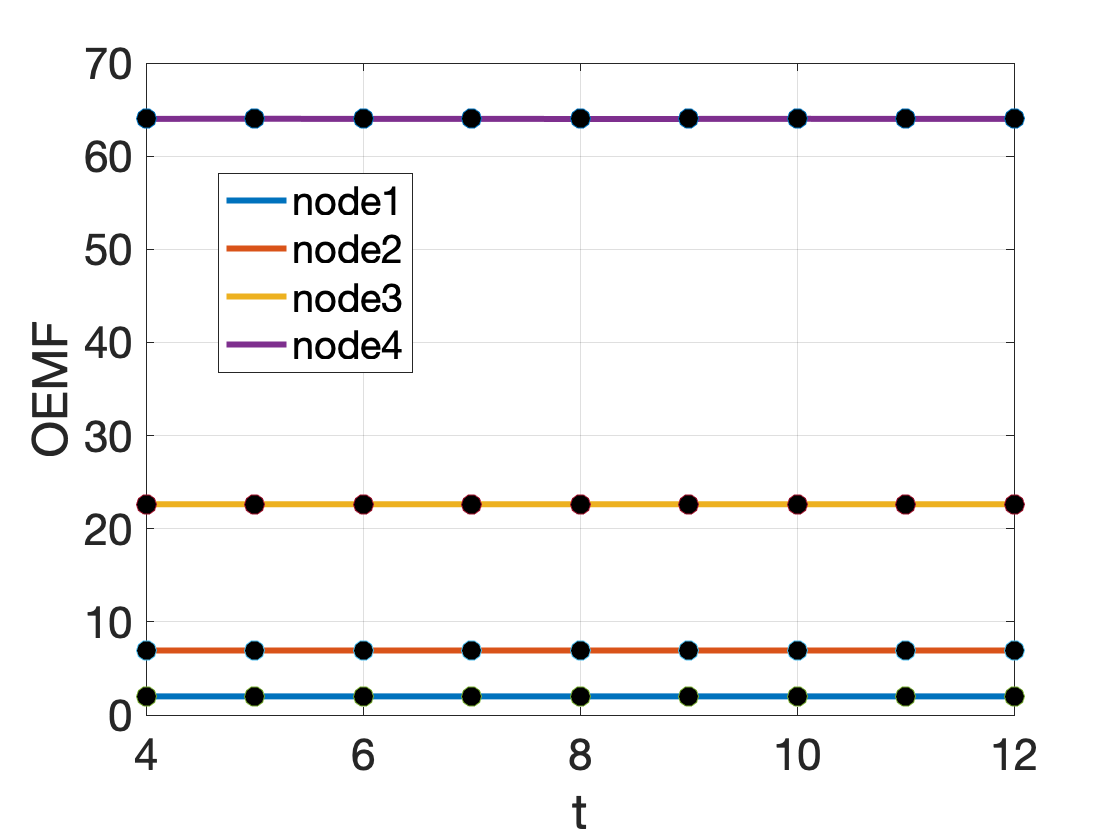}}
 \subfigure[]{ \includegraphics[width=.47\textwidth]{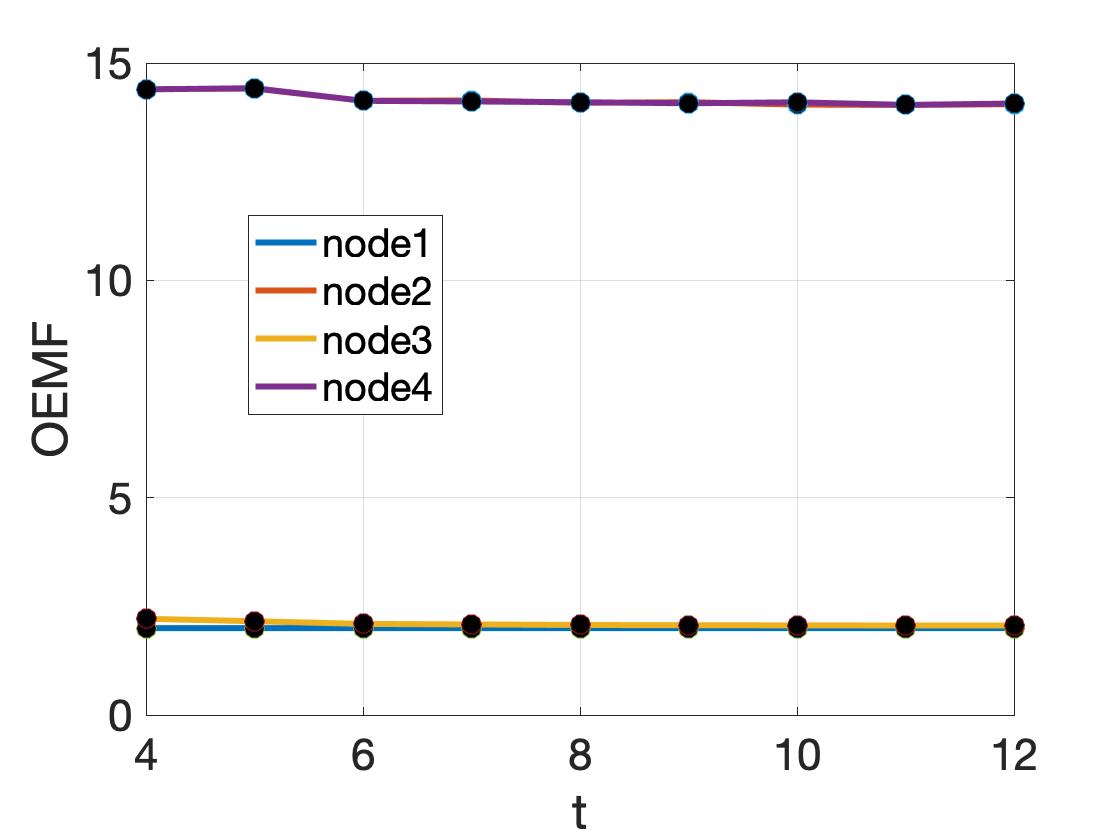}}
\end{center}
\caption{\rm OEMF for 4-node network in Fig.~\ref{f1}(a).   (a) The OEMF $\kappa^t_{1i}$ for $i = 1,\ldots, 4$. The vertical axis displays the noise magnification when reconstructing nodes from the time series at node 1. The OEMF appear to tend to a limit as the trajectory length $t$ increases. (b) The OEMF  $\kappa^t_{2i}$ for each $i$. The vertical axis displays the noise magnification when reconstructing nodes from the time series at node 2.  In both cases, the lowest OEMFs occur for the diagonally opposed node, not the ones directly connected to the observer. } \label{f4}
\end{figure}

It follows from Lemma \ref{lm2} that the RMS of the perturbations in the reconstruction at node $i$ is given by
\begin{equation}\label{ek}
 {\rm RMS}(h) = \frac{1}{\sqrt{t}} ||M_{t,i} M_{t,j}^\dagger ||_F\ {\rm RMS}(e).
 \end{equation}
Thus stepwise noise of size $\sigma$ inserted to the length $t$ time series at the observation node $j$ will result in magnification of the reconstructed time series by
\begin{equation}\label{e9}
\frac{{\rm RMS}(h)}{{\rm RMS}(e)} = \frac{1}{\sqrt{t}} \kappa_{ji}^t
 \end{equation}
where we have defined $\kappa_{ji}^t  \equiv ||M_{t,i}\ M_{t,j}^\dagger||_F$.
  
 \begin{figure}
 \begin{center}
 \subfigure[]{ \includegraphics[width=.45\textwidth]{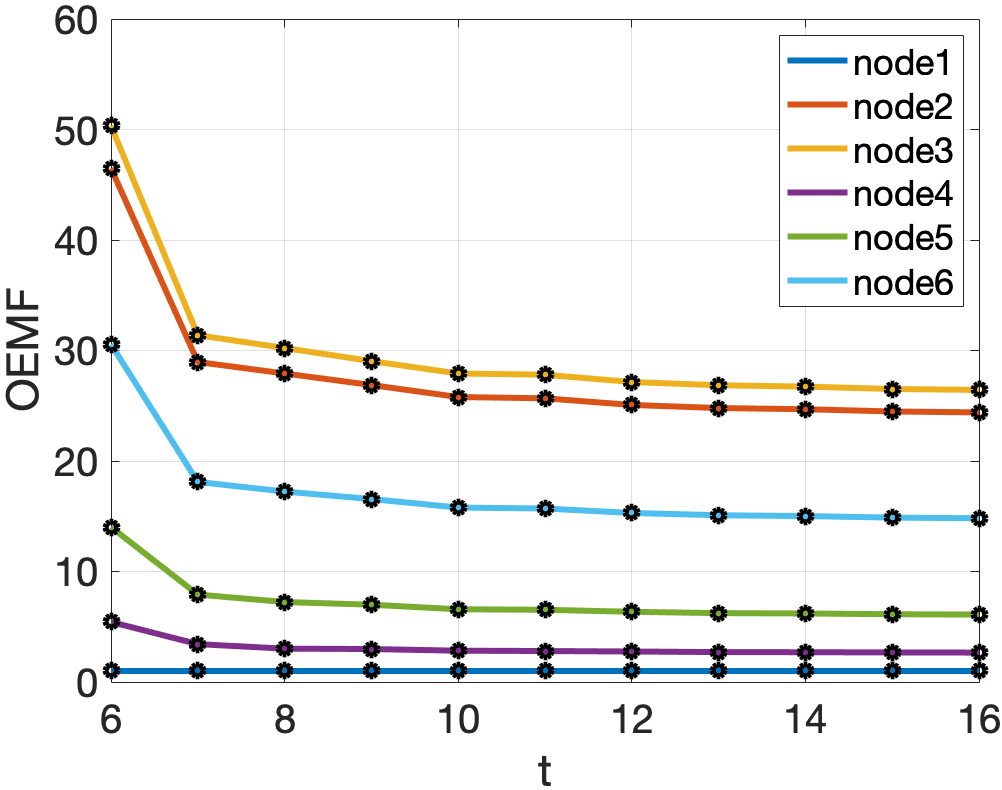}}
\subfigure[]{ \includegraphics[width=.45\textwidth]{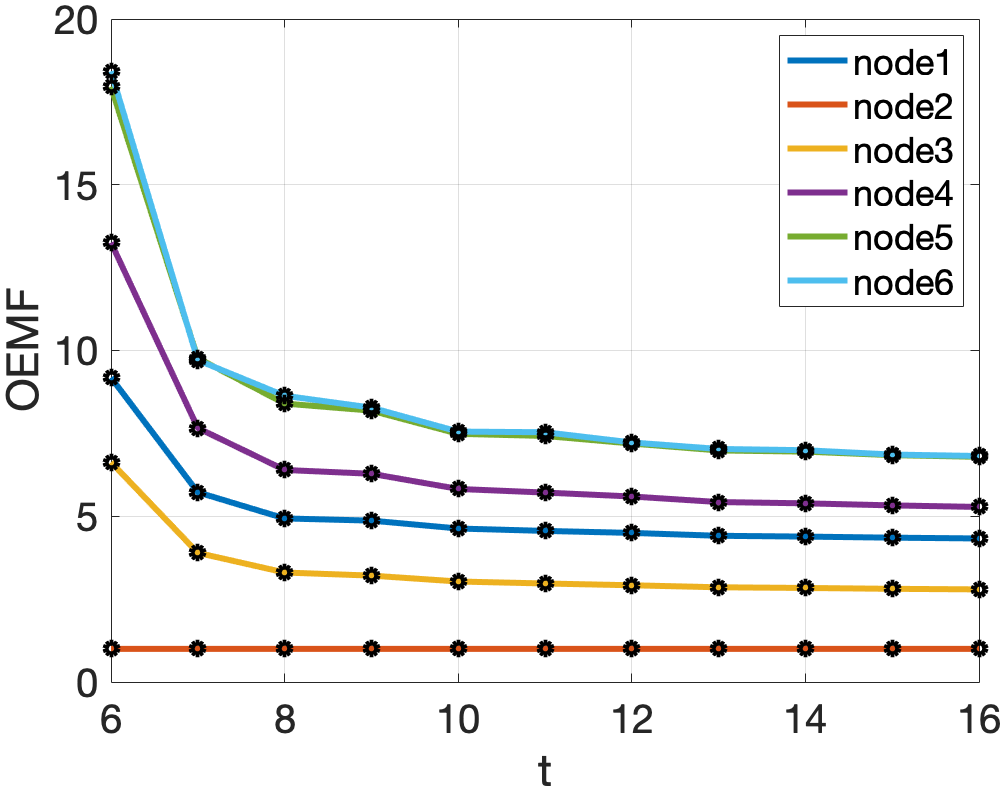}}
\end{center}
\caption{\rm OEMF for 6-node network in Fig.~\ref{f2}(b). (a) The OEMF $\kappa^t_{1i}$ for $i = 1,\ldots, 6$. The OEMF  with respect to the observing node $1$ tend to a limit as the trajectory length $t$ increases. The reason that the OEMFs are higher for reconstructing nodes $2$ and $3$ is not obvious from the network topology. (b) The OEMF  $\kappa^t_{2i}$ for each $i$, with respect to the observing node $2$. Nodes $5$ and $6$ are the most difficult to reconstruct from node $2$. While the curves are not as flat as in Fig.~\ref{f4}, they appear to asymptote to a constant value as $t\to\infty$, consistent with definition (\ref{e20}). } \label{f5}
\end{figure}

As a consistency check, in the special case when we reconstruct the $j$-th node from observations of itself, and $M_{t,j}$ is full rank $n$, then $$M_{t,j}M_{t,j}^\dagger  = USV^TVS_{\rm inv}U^T = USS_{\rm inv}U^T = U(1:n)U(1:n)^T $$ where $U(1:n)$ denotes the first $n$ columns of $U$. In this case,
\begin{eqnarray*}
 \kappa_{jj}^t & =& ||M_{t,j}M_{t,j}^\dagger||_F =  \sqrt{{\rm tr}\ (U(1:n)U(1:n)^T)^T U(1:n)U(1:n)^T} \\&=&  \sqrt{{\rm tr}\ U(1:n)U(1:n)^T} = \sqrt{{\rm tr}\ U(1:n)^TU(1:n)} = \sqrt{n}, 
 \end{eqnarray*} and (\ref{e9}) shows that 
\[ {\rm RMS}(h) = \frac{1}{\sqrt{t}}\ \kappa_{jj}^t\ {\rm RMS}(e) = \frac{\sqrt{n}}{\sqrt{t}}{\rm RMS}(e). \]
 For $t=n$, ${\rm RMS}(h) = {\rm RMS}(e)$ as expected. 
 For $t>n$, $ {\rm RMS}(h) <  {\rm RMS}(e)$ because the longer time series provides more information, in a similar way to the general fact that taking more samples improves the estimate of a mean. 
 
 An alternative way of calculating $\kappa_{ji}^t$ is through the $QR$-factorizations of $M_{t,i} = Q_{t,i}R_{t,i}$ and $ M_{t,j}=Q_{t,j}R_{t,j}$.  Then 
 $$\kappa_{ji}^t = ||R_{t,i}(1:n)R_{t,j}(1:n)^\dagger||_F
$$ where $R_{t,i}(1:n)$ denotes the first $n$ rows of $R_{t,i}$. The advantage of this formulation is that the Frobenius norm is taken over an $n\times n$ matrix for all $t$.

 What happens to $\kappa_{ji}^t$ as the length $t$ of the trajectory increases? On the one hand, we expect the error magnification to decrease since there is more information in longer time series. On the other hand, the Frobenius norm in (\ref{e9}) is taken over a matrix of increasing size $t\times n$. 
We conjecture that in the case of distinct $i$ and $j$ , the limit of $\kappa_{ji}^t$ reaches a limit, which we denote $\kappa_{ji}$ as  $t\to\infty$.  We examine this question in the  following three examples.

\begin{examp}\label{ex5}\rm
Consider the 4-node network sketched in Fig.~\ref{f1}(a). We define a weight matrix respecting the graph, with nonzero weights chosen as $a_{ij} = 1+0.5\nu_{ij}$, where $\nu_{ij} \in N(0,1)$, and consider the discrete linear dynamical system produced. Fig.~\ref{f4} shows the mean $\kappa_{ij}^t$ taken over $10^4$ realizations of the weight matrices. for (a) $j=1$ (time series measured at node 1) and (b) $j=2$ (time series measured at node 2). 

We note two interesting observations from the result. First, the value of each $\kappa_{ji}^t$ is not constant with $t$, and appears to monotonically decrease with $t$ to a limiting value as proposed in the above definition. Second, one may not have guessed the relative sizes of the $\kappa_{ji}^t$ from the weighted graph in Fig.~\ref{f1}(a) that defines the dynamics. For the observations at node 1 in Fig.~\ref{f4}(a), the ``farthest'' node 3 from node 1 is has the least error magnification, and the nodes adjacent to node 1 have the largest error magnification. Likewise in Fig.~\ref{f4}(b), it is node 3, which is not directly connected to node 2, which has by far the lowest error magnification. This apparently shows the utility of the $\kappa_{ji}^t$ to identify the practicality of reconstruction, that may not be obvious by other means.
\end{examp}

\begin{examp}\label{ex6}\rm
In the six-node network of Fig.~\ref{f2}(b), we added an edge from node 5 to node 6, to destroy the bottleneck and guarantee reconstruction from observations at nodes 1 or 2. As in the previous example, we establish weights as $a_{ij} = 1+0.5\nu_{ij}$ corresponding to each directed edge in the graph, where $\nu_{ij} \in N(0,1)$, and average $\kappa_{ij}^t$ over $10^4$ realizations of the weights. Fig.~\ref{f5} shows the results.
\end{examp}

\begin{figure}
 \begin{center}
 \subfigure[]{ \includegraphics[width=.45\textwidth]{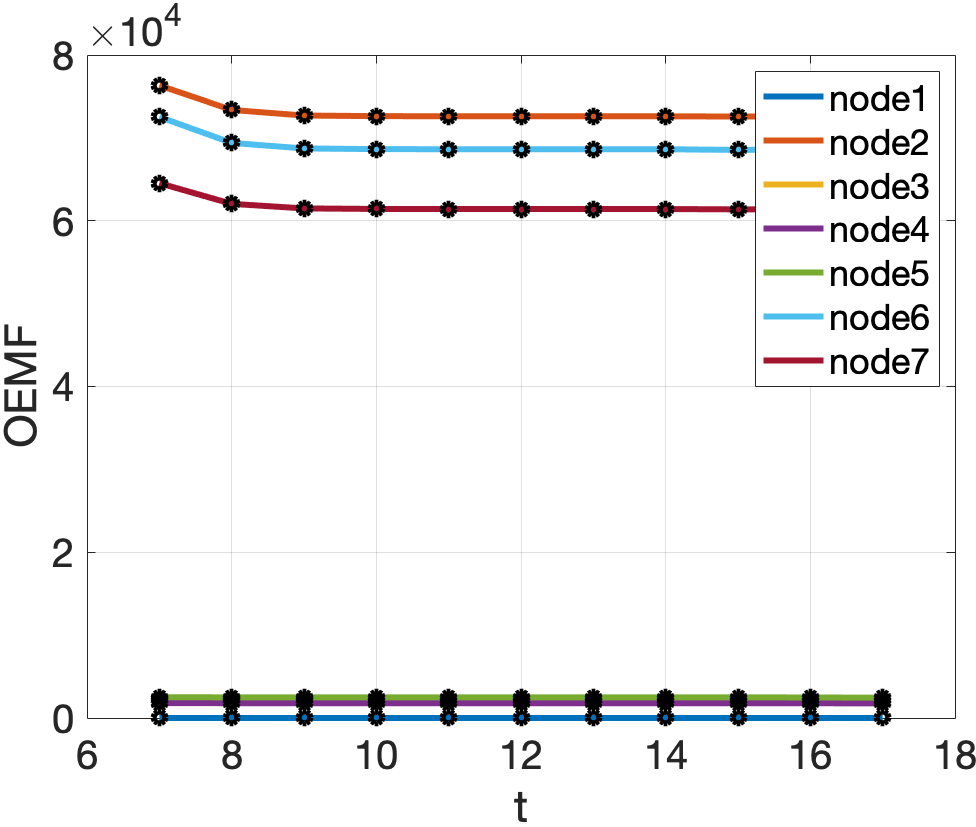}}
\subfigure[]{ \includegraphics[width=.45\textwidth]{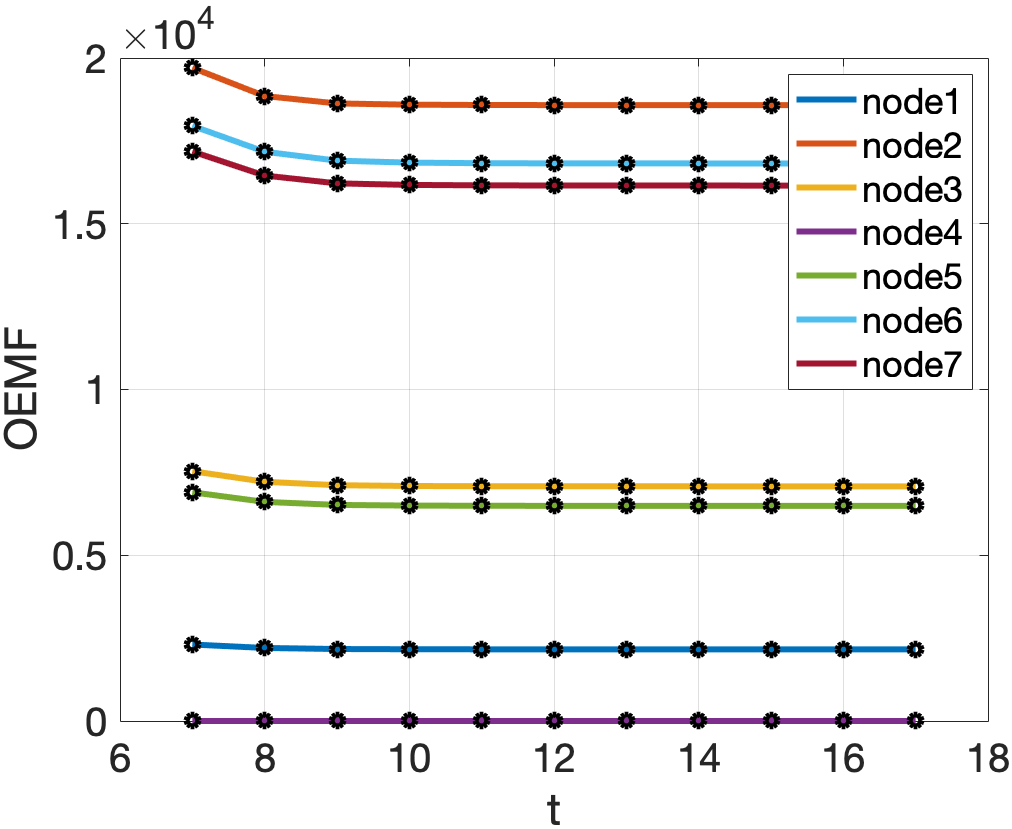}}
\end{center}
\caption{OEMF for the 7-node network in Fig.~\ref{f3}(b). (a)  $\kappa^t_{1i}$ from observations at node 1. Nodes 2, 6, and 7 are significantly more difficult to reconstruct from node 1. (b) $\kappa^t_{4i}$ from observations at node 4. The same three most difficult nodes persist.} \label{f6}
\end{figure}

\begin{examp}\label{ex7} \rm  The average $\kappa^t_{i1}$ and $\kappa^t_{i4}$ for the system of Fig.~\ref{f3}(b)  are shown in Fig.~\ref{f6}. Recall that there was an obstruction to reconstructibility in Fig.~\ref{f3}(a), which was relieved by adding one extra directed edge. As in the above examples,  we generate weights as $a_{ij} = 1+0.5\nu_{ij}$ corresponding to each directed edge in the graph, where $\nu_{ij} \in N(0,1)$, and average $\kappa_{ji}^t$ over $10^4$ realizations of the weights. The system has relatively high error magnification, perhaps due to being near the border of reconstructibility.

\end{examp}

In Examples \ref{ex5}, \ref{ex6}, and \ref{ex7}, it is apparent that the $\kappa_{ji}^t$ are monotonically decreasing in the trajectory length $t$, and that they appear to approach a limit.
These examples motivate the following definition, which considers the limit of (\ref{e9}) as $t\to\infty$, if it exists.

\bigskip

\noindent{\bf Defn.} \cite{guan2018limits} Let $[x^0,\ldots, x^t]$  be a trajectory at node $X$, and let $[x^0+h^0,\ldots, x^t+h^t]$ be a trajectory reconstructed from observations on the set $S$. The {\bf observational error magnification factor (OEMF)} of the trajectory is defined to be
\begin{eqnarray} \label{e20}
\kappa_{S,X} &\equiv& \lim_{\sigma\to 0}\ \lim_{t\to\infty}\frac{\left\{\mathbb{E} \left[ ||h||_2^2\right]\right\}^{1/2}}{\sigma}\nonumber\\
&=& \lim_{\sigma\to 0}\ \lim_{t\to\infty}\frac{\sqrt{t}\ \mathbb{E} \left[ ||h||_2^2/t\right]^{1/2}}{\sigma}\nonumber\\
&\approx&  \sqrt{t}\ \frac{\rm RMS\ reconstruction\ error\ per\ step\ at\ X}{\rm RMS\ observation\ error\ per\ step\ at\ S}
 \end{eqnarray}
 As we showed above, in the discrete linear case with $S = \{j\}$ and $X=\{i\}$, $\kappa_{S,X} = \kappa_{ji}$.
 
ln the general nonlinear case, we can also expect a constant $\kappa_{S,X}$ that is independent of the length of the trajectory and the size of the observational noise, at least in the limiting case. The significance of this definition is that it is useful to have a single number which characterizes the ability to reconstruct a time series at a node $X$ from observations at a subset $S$ of the network. We will pursue this larger context in the next two sections.

\section{Numerical method}

In this section we describe a numerical algorithm for reconstructing time series at a network node from observations at a downstream node. In the previous section, we accomplished this in the discrete linear case, when such a reconstruction was possible. In this section we pursue the same question for nonlinear dynamics.

Since we know the dynamical equations, the problem is closely related to reproducing an initial condition of the entire system that generates the trajectory. We will assume that we can observe the entire time series at a subset $S$ of network nodes, including the initial condition at those nodes, but have no such knowledge at the desired node $X$.

Our computational approach will consist of minimizing a loss function on potential full trajectories, that simultaneously monitors both the distance from the observations and the discrepancy of the trajectory's iterations from exactness. We note that whether the observations are made with or without noise, this minimization is highly nonconvex, and we will find that calculating a minimum is often nontrivial.

\subsection{Loss function} \label{ssec31}
Let $f:R^n\to R^n$ be a discrete map, and let $\{x_i^k\}$ be an exact trajectory of $f$ for $1 \leq i \leq n$ and $0\leq k \leq t$.
Assume that $m\geq 1$ nodes are observed, and renumber them as nodes $1, \dots, m$ for simplicity. Up to this point, we have shown examples with $m=1$, but we allow $m>1$ in general for the case where more than one node can be observed.
In the absence of noise,  the following $m(t+1)+nt$  equations in $n(t+1)$ unknowns  $x_i^k$ hold for the trajectory:
\begin{eqnarray} \label{eq}
x_i^k&=& s_i^k  {\rm\ \ \ \ \ \ \ \ \  \ \ \  for\ \ } i=1,\ldots, m {\rm \ \ and\ \ } k=0,\ldots,t \\
x_i^{k+1}& =& f(x_i^k) {\rm\ \ \ \ \ \ \ \ \ for\ \ } i=1,\ldots, n {\rm \ \ and\ \ } k=0,\ldots,t-1 .  \nonumber
\end{eqnarray}
In a realistic application, the nodes are observed with noise, i.e. $$s_i^k = x_i^k + e_i^k$$
for $i = 1,\ldots, m$. 
Our goal is to reconstruct $x_i^k$ for $i = 1, \ldots, n$ and $k = 0, \ldots, t$ only from knowledge of the observations $s_i^t$ for $i=1, \ldots, m$ and the dynamical equations denoted by $f$.
Let $\{y_i^k\}$ be the target time series and denote $h_i^k = y_i^k - x_i^k$ the errors in reconstructing the exact trajectory.

\begin{figure}
 \begin{center}
 \subfigure[]{ \includegraphics[width=.48\textwidth]{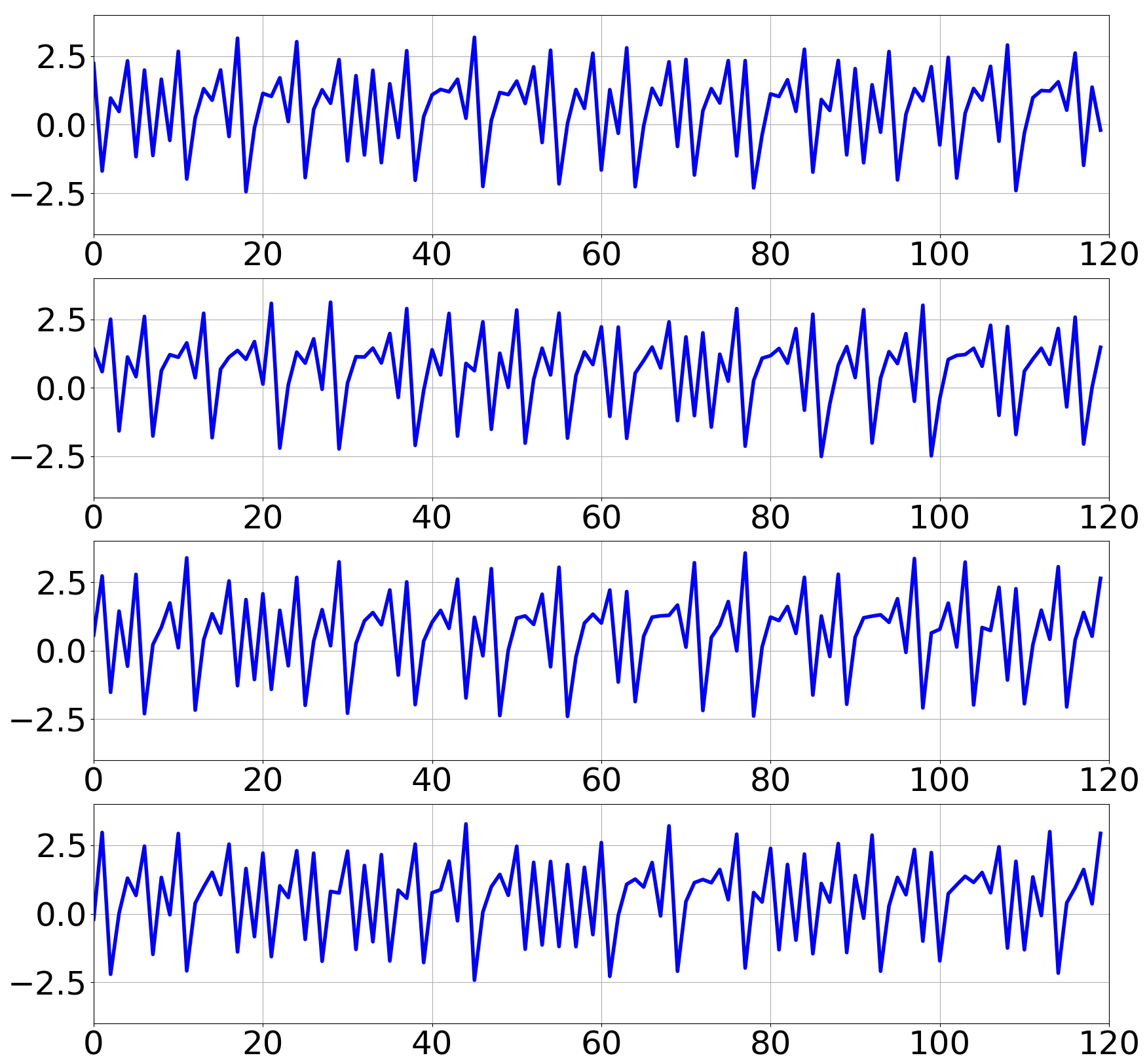}}
\subfigure[]{ \includegraphics[width=.48\textwidth]{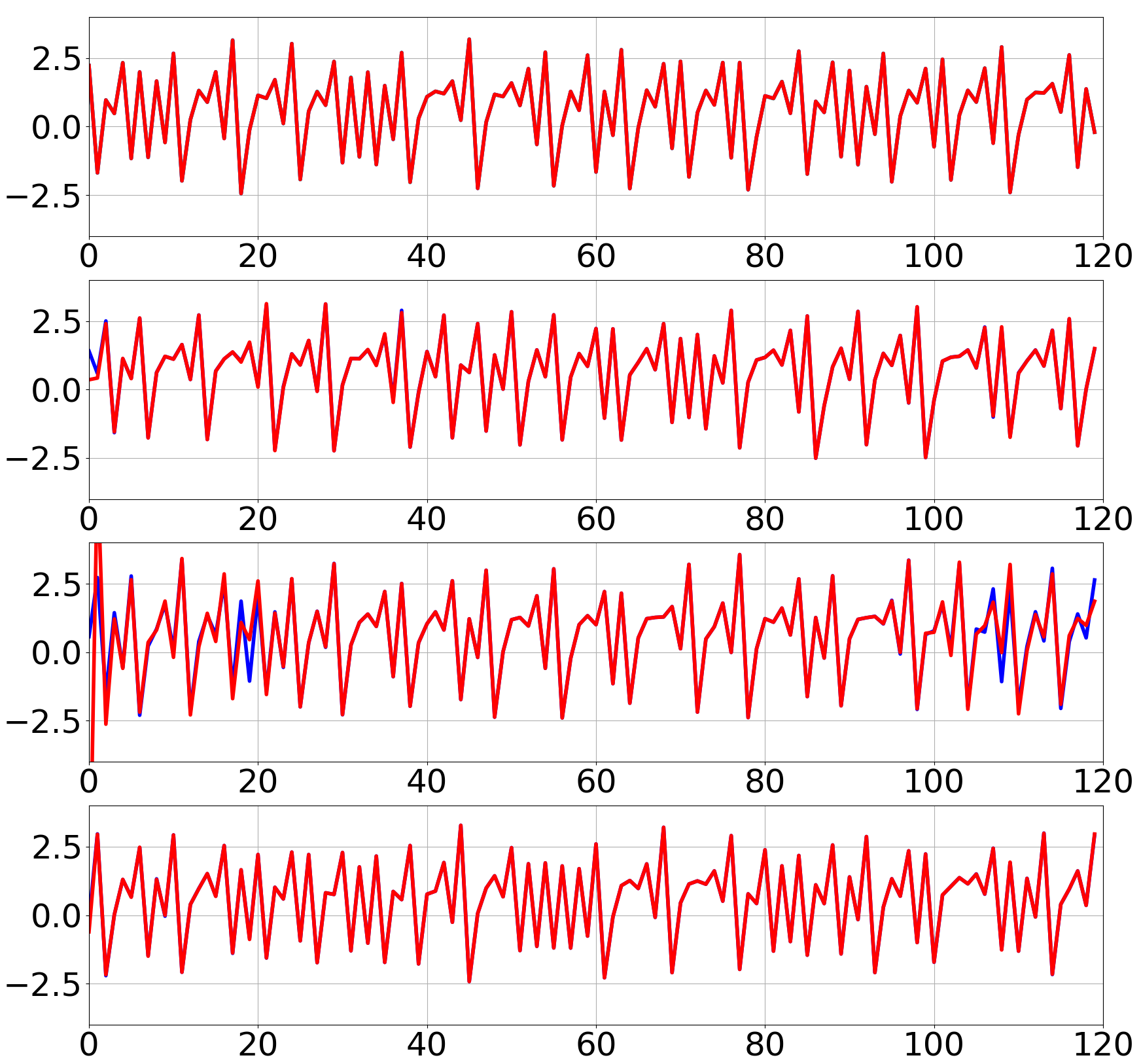}}
\end{center}
\caption{\rm Reconstruction of times series in a 4-node network with topology as in Fig.~\ref{f1}(b) with local H\'enon dynamics (\ref{eh}) at each node. (a) An exact trajectory (in blue) of length 120, plotted at each of the four nodes; nodes arranged vertically by number. (b) Reconstruction (in red) produced by Gauss-Newton  from  observations at node 1, observed with noise level $\sigma = 0.03$. The exact trajectories are in plotted in blue. The largest errors are seen in the reconstruction of node 3, consistent with the OEMF displayed in Fig.~\ref{f7}(b). } \label{f7}
\end{figure}

Comparing the number of equations above that the $y_i^k$ must satisfy shows that the equations in (\ref{eq}) are overdetermined as long as the trajectory length $t > (n-m)/m$. For moderately long times series, this requirement will be easy to achieve. Due to the noise, the  $x_i^k$ will not satisfy the equations (\ref{eq}), but we will search for the best least squares alternative $y_i^k$.

The overdetermined  least squares problem that arises is to minimize the loss function 
\begin{equation}\label{eloss}
L_w(y) \equiv w\sum_{k=0}^t \sum_{i=1}^m [y_i^k - s_i^k]^2   +  \sum_{k=0}^{t-1} \sum_{i=1}^n [y_i^{k+1} - f(y_i^k)]^2 
\end{equation}
for a weight $w$.
The first double sum represents the observational discrepancy, the difference between the $y$ trajectory and the noisy observations of the $x$ trajectory. The second double sum represents the consistency discrepancy of the $y$ trajectory, a measurement of how far the $y$ time series is from being an exact trajectory of the dynamical map $f$.

\subsection{Gauss-Newton with QR}

Let $r_1(x), \ldots,r_m(x)$ be functions $r_i: R^n\to R$. To minimize $\sum_{i=1}^m r_i(x)^2$, start with initial guess $x^0$.
Set $r = [r_1(x), \ldots,r_m(x)]$ and denote by $Dr(x)$ the $m\times n$ matrix of partial derivatives. The Gauss-Newton method \cite{sauer} produces the iterates $x^{k+1} = x^k + v^k$ where $v^k$ is the linear least squares solution of $$Dr(x^k) v^k = -r(x^k).$$ 
If $x^0$ is close enough to the optimum, the iterates will converge to it.

In cases where the minimization problem is poorly conditioned, it is helpful to use the $QR$ factorization to compute $v^k$. That is, set $Q_kR_k = Dr(x^k)$ where $Q_k$ is an orthogonal $m\times m$ matrix and $R_k$ is $m\times n$ upper triangular.  Then 
\begin{equation}\label{eGN}
v^k = (R_k^{\rm top})^{-1} Q_k^L \ Dr(x^k)
\end{equation}
similar to the calculation in section \ref{sec3}.

As a local optimization algorithm, the Gauss-Newton method is not guaranteed to converge to the minimum of a nonconvex optimization problem, especially if the initial guess is far from the optimum. In our case, we assume little information about the optimum,  the exact full trajectory, is available.

\begin{figure}
 \begin{center}
 \subfigure[]{ \includegraphics[width=.45\textwidth]{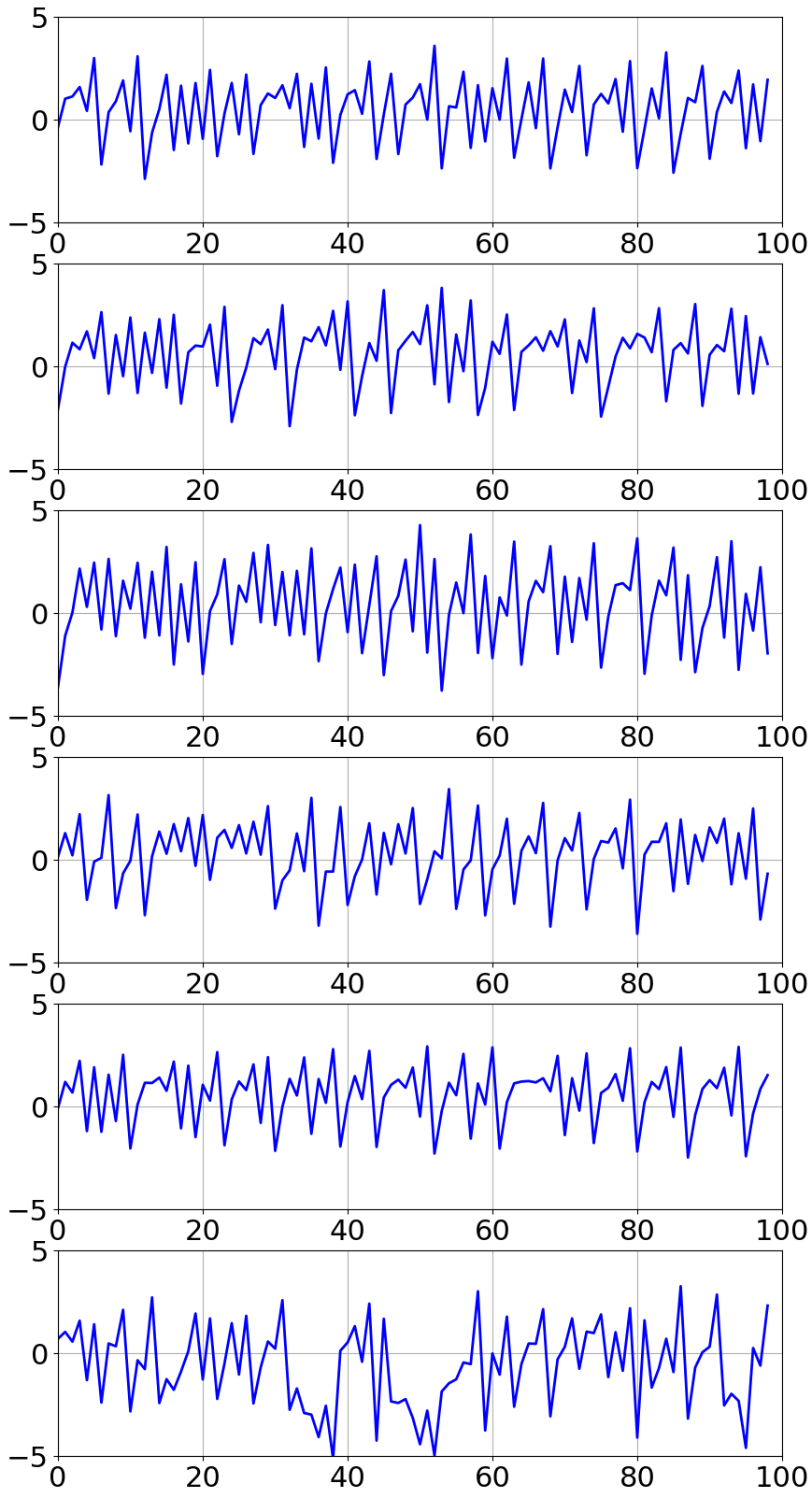}}
\subfigure[]{ \includegraphics[width=.45\textwidth]{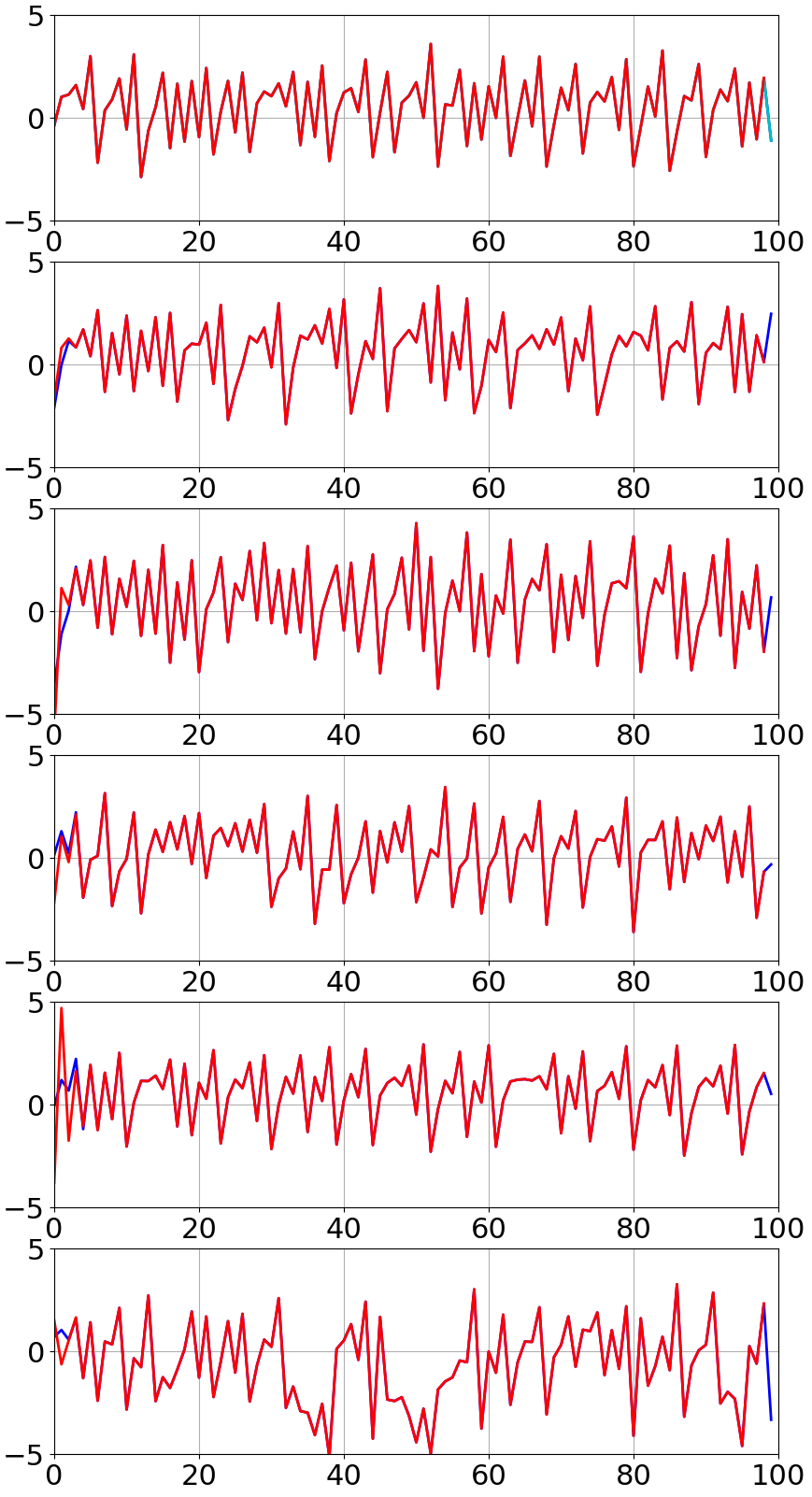}}
\end{center}
\caption{\rm Reconstruction of times series in a 6-node network with topology as in Fig.~\ref{f2}(b) with local H\'enon dynamics (\ref{eh}) at each node. (a) An exact trajectory (in blue) of length 100, plotted at each of the six nodes. (b) Reconstruction (in red)  from  observations at node 1, observed with noise level $\sigma = 0.001$.} \label{f8}
\end{figure}

\subsection{Networks of nonlinear systems}
Most of the computer simulations in this section use network topologies studied above in the discrete linear case, but where each network node has been replaced by a nonlinear map. The examples typically use a modification of the H\'enon map \cite{henon2004two} due to its relative simplicity.

Define the modified H\'enon map $F:\mathbb{R}^2\to\mathbb{R}^2$ by 
\begin{equation*}
F(x,y) = (b\cos x + cy,x).
\end{equation*}
In the following examples, the parameters $b = 2.2, c = 0.4$ are used, and if used in a network, parameter values generated near those values with small changes chosen from a normal distribution are used, to avoid unwanted symmetries. These parameter values results in chaotic dynamics for each node separately, and depending on the influences from other nodes from the network weights, tend to result in chaotic network dynamics.

We will use dynamics on a $n$-node network to illustrate the properties of the error magnification discussed above. Consider the H\'enon-like map $f:\mathbb{R}^{2n}\to\mathbb{R}^{2n}$ defined by
\begin{eqnarray} \label{eh}
x_i^{k+1} &=& b_i\cos x_i^k + c_ix_{i+1}^k + \sum_{j=1}^n a_{ij}x_{2j-1}^k \\
x_{i+1}^{k+1} &=& x_i^k \nonumber
\end{eqnarray}
for $i = 1, 3, 5, \ldots, 2n-1$, where the $a_{ij}$ form an $n\times n$ weight matrix $A$. Thus the network communicates through the odd-numbered variables, one per node, while the even numbered variables are considered ``internal'' or recovery variables. In the following examples, certain of the $a_{ij}$ are fixed at zero to respect a particular directed graph, and the nonzero $b_i, c_i$ are chosen to be normal random perturbations of  $2.2$ and $0.4$, respectively.

\begin{examp} \rm
Fig.~\ref{f7} shows the results of reconstruction of times series in the 4-node network of Fig.~\ref{f1}(a) with observations from the $x$-variable at node 1 (variable $x_1$ in terms of equation (\ref{eh})), the top trace in each of the panels. A small amount of observational noise is added. Panel (a) shows the exact traces, and panel (b) shows the reconstructed traces from the Gauss-Newton iteration plotted in red.  Some deviations from the true trajectory are noticeable at node 3. The fact that observational error is magnified more for the reconstruction of node 3 reflects the predictions of the OEMF as displayed in Fig.~\ref{f10}.\end{examp}

\begin{examp} \rm
In Fig.~\ref{f8}, the reconstruction of a network of H\'enon maps connected as the 6-node network in Fig.~\ref{f2}(b) is carried out. The exact trajectories are shown in panel (a). As in the previous example, only the $x$-variable at the first node is observed, and the remaining 11 traces are reconstructed by the Gauss-Newton method described above and displayed in panel (b).
\end{examp}

\subsection{Reconstruction algorithm}

In this subsection we collect some details on the application of Gauss-Newton to minimize the loss function (\ref{eloss}). As an initial guess for the minimization, we use a short trajectory consisting of the $m$ observed coordinates, observed with noise, and the other $n-m$ coordinates seeded with normal random numbers.

\begin{figure}
 \begin{center}
 \subfigure[]{ \includegraphics[width=.45\textwidth]{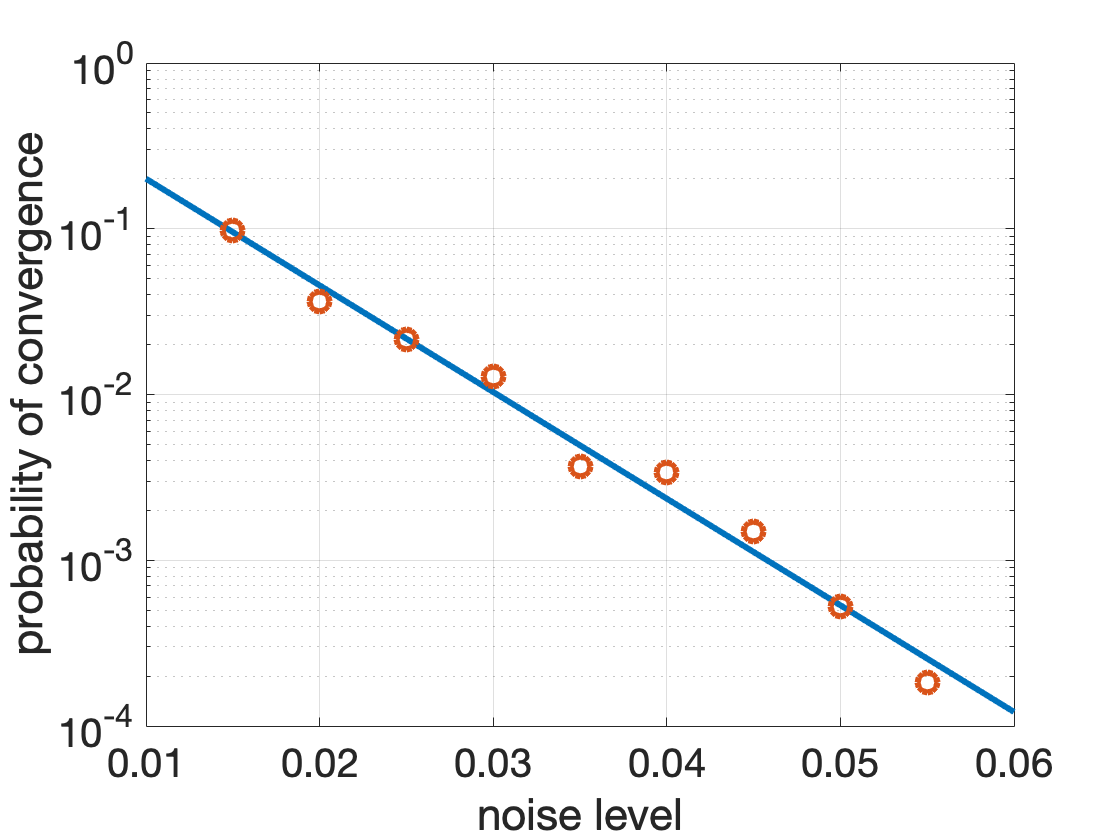}}
\subfigure[]{ \includegraphics[width=.45\textwidth]{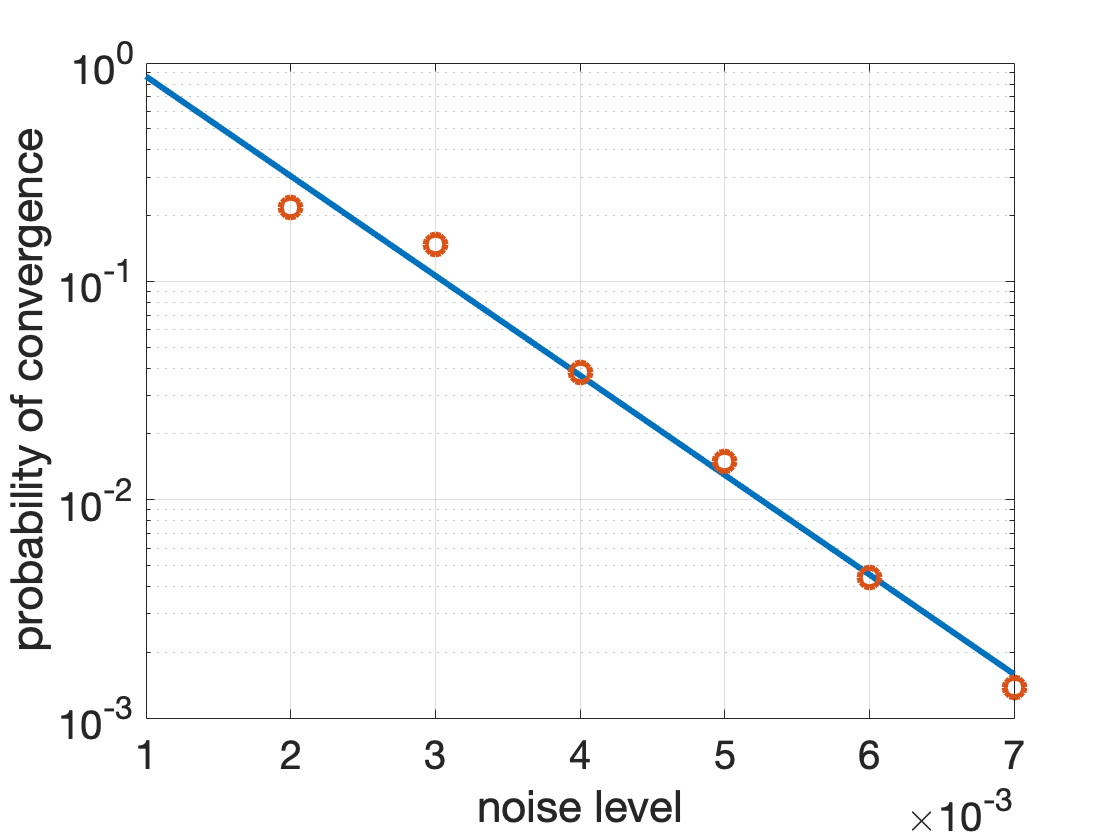}}
\end{center}
\caption{\rm Success curve for damped Gauss-Newton algorithm. For larger noise levels, most initial starts end by diverging to infinity, and only a small fraction converge to a trajectory near the true one. The probability of a convergence is plotted as a function of observation noise level for  (a) the 4-node H\'enon network with the topology of Fig.~\ref{f1}(a), and (b) 6-node H\'enon network with the topology of Fig.~\ref{f2}(b).} \label{f9}
\end{figure}

Starting with random coordinates in this way is a challenge for a local method like Gauss-Newton, which tends to diverge to infinity if it is too far from the minimum. This problem is more prevalent with increasing observation noise level $\sigma$. We address this problem in two ways:

(1) The application of Gauss-Newton algorithm is done with a reduced step size, using the idea which is often called damped Gauss-Newton. In other words, we routinely multiply the proposed Gauss-Newton innovation (\ref{eGN}) by a small number (such as $10^{-p}$ for $p = 1, 2$ or $3$) which can allow the method, once it acquires the basin of convergence, to avoid jumping out of the basin.

(2) Multiple restarts of the damped Gauss-Newton method are needed in most circumstances. For larger noises, hundreds or thousands of restarts (reseeding the initial guess of the trajectory with random numbers) were required to converge to a trajectory close to the original exact trajectory. The growth of the number of restarts needed for convergence, as a function of observation noise level, is analyzed in Fig.~\ref{f9}.

 Fig.~\ref{f9} summarizes important facts about our ability to reconstruct from partial observations. The basins of convergence of damped Gauss-Newton for this problem are extremely complex. It is common to see trajectories that track closely to the desired trajectory for a finite number of steps, and then suddenly diverge from that trajectory. In such a case, the Gauss-Newton has to be reinitialized with a new random start trajectory. The difficulty of finding the correct basin of convergence appears to increase exponentially with the observational noise level, according to the fit shown in the figure. We do not have a theoretical explanation for this scaling.

\section{OEMF for nonlinear networks}
In this section, we use the definition (\ref{e20}) of OEMF derived in section 3 and the numerical method developed in section 4 to estimate the error magnification inherent in some example networks. We rely primarily on the networks of modified H\'enon maps defined above, and use some of the same network topologies from the discrete linear examples in section 2. We will see that the same network topologies exhibit quite different OEMFs in the nonlinear case.

\begin{figure}
\begin{center}
\subfigure[]{ \includegraphics[width=.48\textwidth]{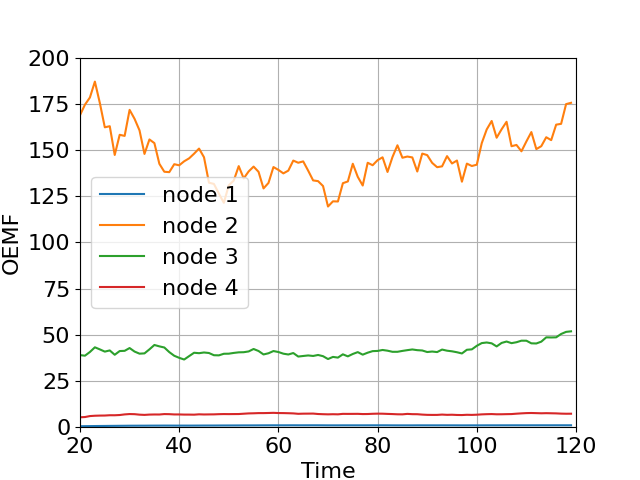}}
\subfigure[]{ \includegraphics[width=.48\textwidth]{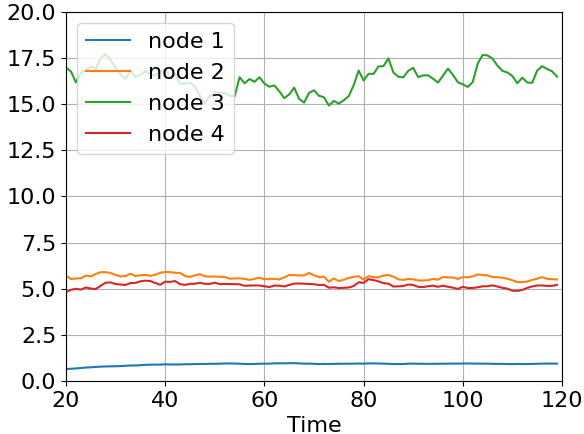}}
\end{center}
\caption{\rm (a) Estimated OEMF of modified H\'enon dynamics connected in a directed circular network as in Fig.~\ref{f1}(a) with observations at node 1 and noise level $10^{-4}$. (b) Estimated OEMF for the undirected circular network as in Fig.~\ref{f1}(b)} \label{f10}
\end{figure}

\begin{examp} \rm
Fig.~\ref{f10}(a) displays an estimate of OEMF from a directed network with topology as in Fig.~\ref{f1}(a), with four nodes arranged in a circle. Noisy observations are made from the first coordinate at node 1, and the remaining seven time series are reconstructed according to the algorithm in section 4. Then the formula (\ref{e20}) reveals the estimated OEMF as a function of step number $t$. Several trajectories of the same system are averaged and plotted versus time.

One notes that the relative sizes of the OEMF are arranged in the same order as the relationship to the observing node 1. In fact, node 4, which directly feeds node 1, has the lowest error magnification, followed by node 3 and node 2, which has the longest path to node 1. 

Fig.~\ref{f10}(b) shows the results from an undirected network with circular topology, as depicted in Fig.~\ref{f1}(b). As before, noisy observations are made from node 1. In this case, the OEMF at nodes 2 and 4 are essentially the same, as can be expected. The highest OEMF occurs for node 3, which is farthest in path distance from node 1. However, all error magnification factors are lower than in the corresponding directed network case in panel (a) of the figure.

\end{examp}

\begin{figure}
 \begin{center}
  \subfigure[]{ \includegraphics[width=.45\textwidth]{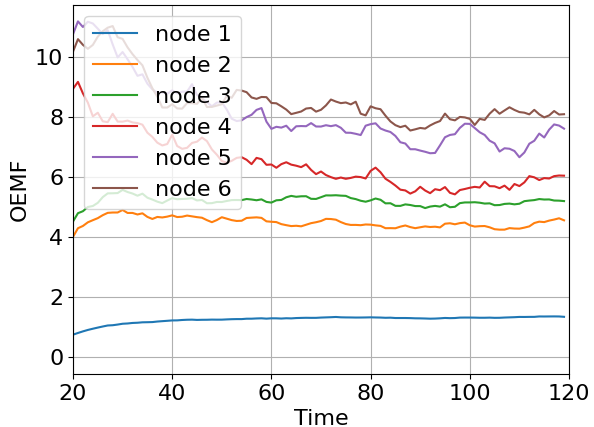}}
\subfigure[]{ \includegraphics[width=.45\textwidth]{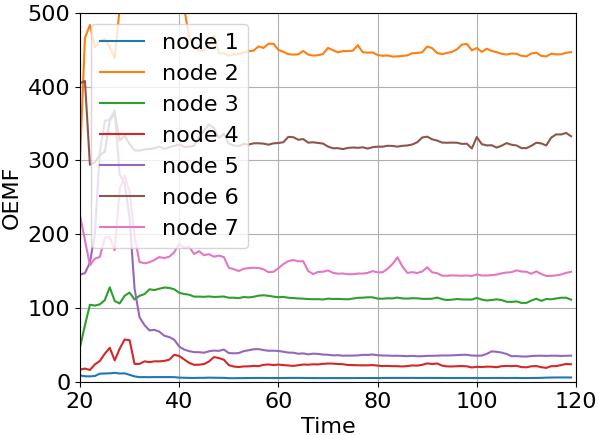}}
\end{center}
\caption{\rm (a) Estimated OEMF for 6-node network of modified Henon maps arranged according to Fig.~ \ref{f2}(b). (b) Estimated OEMF for 7-node network arranged according to Fig.~\ref{f3}(b). Comparing to Figs.~\ref{f5} and \ref{f6}, the nonlinear versions show the same difference in reconstruction difficulty between panels (a) and (b) as the discrete linear case.} \label{f11}
\end{figure}

\begin{examp} \rm 
Fig.~\ref{f11}(a) shows the  OEMF for the six-node network of Fig.~\ref{f2}(b). The size of the OEMF are increasing consecutively from node 1 to node 6. It is not very clear from the network topology why this is the correct order, which is one reason that our ability to easily estimate the OEMF from simulation is useful.
\end{examp}

\begin{examp} \rm 
Fig.~\ref{f11}(b) shows the  OEMF for the seven-node network of Fig.~\ref{f3}(b). The relative sizes of OEMF are informative: The middle tier (nodes 3, 4 and 5) are the easiest to reconstruct from node 1, followed by the upper tier (nodes 6 and 7), leaving node 2 to be the most difficult. We have no explanation for this type of effect, and in fact it would be extremely useful to find a way to predict such phenomena from the topology, for example from the characteristics of the allowable paths, etc.
\end{examp}

\begin{figure}
 \begin{center}
  \includegraphics[width=.9\textwidth]{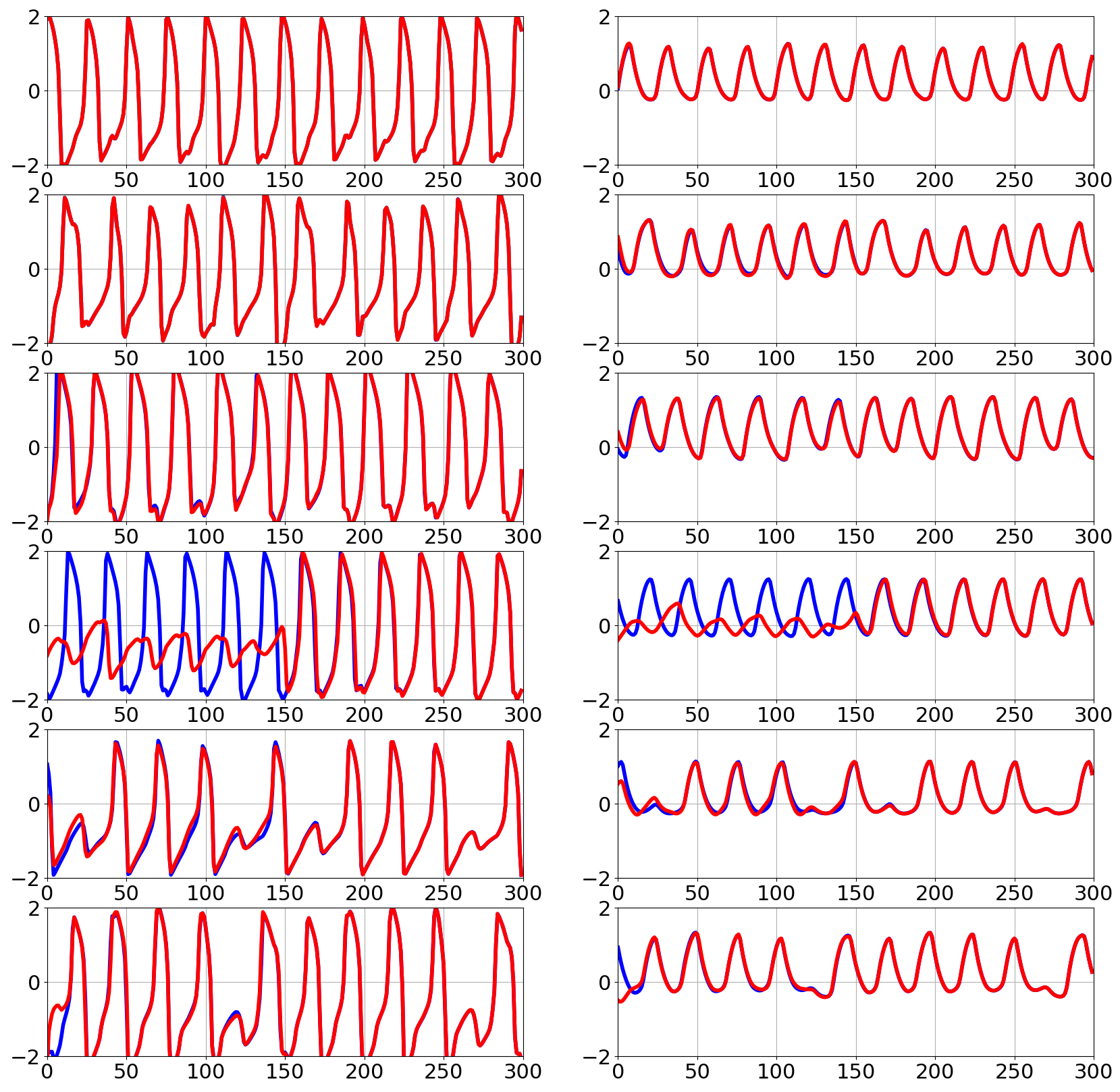}
\end{center}
\caption{\rm Reconstruction of times series in a 6-node network with topology as in Fig.~\ref{f2}(b) with local Fitzhugh-Nagumo  dynamics (\ref{efn}) at each node. Two signals were observed (the first two on top left) and the remaining ten traces were reconstructed from the dynamical equations. Left: An exact trajectory (in blue) of length 300, plotted at each of the six nodes, along with reconstruction (in red)  of the $v$ variables using  observations only  at nodes 1 and w, observed with noise level $\sigma = 0.01$.  Right: The 6 ``recovery variable'' $w$ time series are plotted on the right.} \label{f12}
\end{figure}

\begin{examp} \rm
The method that we propose can also be applied to differential equations. The Fitzhugh-Nagumo neural model \cite{fitzhugh1955mathematical,nagumo1962active} is given by
\begin{eqnarray*}
\dot{v} &=& bv+cw+d - v^3/3\\
\dot{w}&=& ev+fw+g.
\end{eqnarray*}
The parameters are set to be small perturbations of the set $b = 1, c = -1, d = 0.36, e = 0.08, f = -0.06, g = 0.06$. Communication between neurons is established in analogy with (\ref{eh}), by adding to each $v_i$ voltage variable the contributions of all other $v_i$ variables according to the network topology. That is, at each node we solve
\begin{eqnarray*} \label{efn}
\dot{v}_i &=& b_iv_i+c_iw_i+d_i - v_i^3/3 + h \sum_{j=1}^n a_{ij}v_j\\
\dot{w}_i&=& e_iv_i+f_iw_i+g_i
\end{eqnarray*}
where $h = 0.4$.

In order to apply the numerical method described above, we denote by $f$ the time-$\tau$ map of the differential equation, with $\tau = 1$. The derivative of the time $\tau$-map is extracted in the standard way, i.e. by solving the variational equations of the system on each observation step for $\tau$ times units, starting with initial vectors equal to the elementary coordinate vectors. This derivative is updated on each step and used in the application of the Gauss-Newton minimization.

Fig.~\ref{f12} shows the reconstruction of a network of Fitzhugh-Nagumo models. The dynamics are weakly chaotic. The rightmost panel shows the $w$, or recovery, variables. In this example, 10 unobserved traces are successfully reconstructed from the voltage traces (the $v_i$ variable) observed at nodes 1 and 2. Node 4 turns out to be relatively difficult at first, but after about time 150, the reconstruction is quite accurate. It is somewhat interesting that the most periodic trace is the most problematic to extract for this network.
\end{examp}

\section{Discussion} 
The primary goal of this article is to establish that for a trajectory of a dynamical network, there is a single number, the Observational Error Magnification Factor (OEMF), that quantifies the ability to reconstruct the trajectory at a given node by observing at a different node, or subset of nodes, of the network. The OEMF can be used in two obvious ways: (1) to decide how to choose where to extract observations of the network, in order to best monitor the dynamics at another node, or (2) once an observation node is chosen, to quantify how faithful the reconstruction will be at unobserved nodes. The OEMF can be calculated by simulation, in advance of the collection of data, as long as a faithful model of the dynamical network is known.

Our second goal is to propose a plausible numerical algorithm for obtaining the trajectory at unobserved nodes of the network from partial observations elsewhere in the network. We showed results from applying a modified Gauss-Newton iteration to minimize a loss function, which eventually converges to a nearby trajectory, despite potentially requiring a large number of restarts. We were able to quantify the exponential scaling of the number of restarts required.

The exponential scaling of restarts is a reflection of the complicated convergence basin structure of pseudo-trajectories, near the true trajectories of a complex network. Loosely speaking, there are a plethora of trajectories that follow a desired trajectory for a relatively short time and then diverge from it. This is in fact a well-known characteristic of high-dimensional chaotic systems. Shedding light on this fascinating basin structure would be a way to increase the capabilities of this approach.

The application of damped Gauss-Newton should be regarded as only an initial attempt to reconstructing time series from chaotic networks. It is an open question whether a more sophisticated optimization approach could improve the exponential success curves in Fig.~\ref{f8} or perhaps achieve subexponential convergence rates. 

In this article, we have studied the effect of observational noise as a first step. Systems with dynamical noise or model error will present another important source of error magnification and an additional challenge for time series reconstruction, and for analysis of how noise affects reconstruction more generally.

A future application of the ability to quantify error magnification is to predict  the potential success of reconstruction at specific nodes based on the topology of the network and properties of the node dynamics. This is important both for analysis of existing networks and for optimal design of proposed networks. Examples shown here indicate that especially in the nonlinear dynamics case, such predictions may not be straightforward. Our hope is that the availability of easily-computable invariants like the OEMF can be leveraged to investigate these questions.

\newpage
\bibliography{partial}{}
\bibliographystyle{plain}

\end{document}